\documentclass[twocolumn,english]{revtex4}
\usepackage[T1]{fontenc}
\usepackage[latin9]{inputenc}
\usepackage{color}
\usepackage{babel}
\usepackage{float}
\usepackage{amsmath}
\usepackage{amssymb}
\usepackage{stackrel}
\usepackage{graphicx}
\usepackage{esint}
\usepackage[unicode=true,pdfusetitle,
 bookmarks=true,bookmarksnumbered=false,bookmarksopen=false,
 breaklinks=false,pdfborder={0 0 1},backref=false,colorlinks=false]
 {hyperref}
\usepackage{breakurl}

\makeatletter

\newcommand*\LyXThinSpace{\,\hspace{0pt}}

\@ifundefined{textcolor}{}
{%
 \definecolor{BLACK}{gray}{0}
 \definecolor{WHITE}{gray}{1}
 \definecolor{RED}{rgb}{1,0,0}
 \definecolor{GREEN}{rgb}{0,1,0}
 \definecolor{BLUE}{rgb}{0,0,1}
 \definecolor{CYAN}{cmyk}{1,0,0,0}
 \definecolor{MAGENTA}{cmyk}{0,1,0,0}
 \definecolor{YELLOW}{cmyk}{0,0,1,0}
}

\newcommand{\be}{\begin{equation}}\newcommand{\ee}{\end{equation}}\newcommand{\ba}{\begin{array}}\newcommand{\ea}{\end{array}}\newcommand{\bea}{\begin{eqnarray}}\newcommand{\eea}{\end{eqnarray}}

\numberwithin{lemma}{section}\numberwithin{corol}{section}\numberwithin{prop}{section}\numberwithin{dfn}{section}

\usepackage{babel}

\usepackage{babel}

\usepackage{babel}

\usepackage{babel}

\usepackage{babel}

\usepackage{babel}

\usepackage{babel}

\usepackage{afterpage}

\usepackage{babel}

\usepackage{pgfplots}

\usepackage{babel}

\usepackage{babel}

\usepackage{babel}

\usepackage{babel}

\makeatother

\begin{document}

\title{Simple Impedance Response Formulas for the Dispersive Interaction
Rates in the Effective Hamiltonians of Low Anharmonicity Superconducting
Qubits}

\author{Firat Solgun\textsuperscript{1}, David P. DiVincenzo\textsuperscript{2,3}
and Jay M. Gambetta\textsuperscript{1}}

\affiliation{1 IBM T.J. Watson Research Center, Yorktown Heights, NY 10598, USA}

\affiliation{2 Institute for Quantum Information, RWTH Aachen, Germany }

\affiliation{3 Peter Grünberg Institute: Theoretical Nanoelectronics, Research
Center Jülich, Germany}

\date{December 21st, 2017}
\begin{abstract}
For superconducting quantum processors consisting of low anharmonicity
qubits such as transmons we give a complete microwave description
of the system in the qubit subspace. We assume that the qubits are
dispersively coupled to a distributed microwave structure such that
the detunings of the qubits from the internal modes of the microwave
structure are stronger than their couplings. We define ``qubit ports''
across the terminals of the Josephson junctions and ``drive ports''
where transmission lines carrying drive signals reach the chip and
we obtain the multiport impedance response of the linear passive part
of the system between the ports. We then relate interaction parameters
in between qubits and between the qubits and the environment to the
entries of this multiport impedance function: in particular we show
that the exchange coupling rate $J$ between qubits is related in
a simple way to the off-diagonal entry connecting the qubit ports.
Similarly we relate couplings of the qubits to voltage drives and
lossy environment to the entries connecting the qubits and the drive
ports. Our treatment takes into account all the modes (possibly infinite)
that might be present in the distributed electromagnetic structure
and provides an efficient method for the modeling and analysis of
the circuits.
\end{abstract}
\maketitle

\section{Introduction}

Superconducting circuits are a promising platform for the realization
of quantum computers. Operated at microwave frequencies they include
Josephson junctions for the non-linearity needed to obtain qubit modes
without introducing significant loss. Coherence times of the superconducting
qubits have been improved by several orders of magnitude in the last
two decades and the Transmon qubit \cite{Koch-Transmon,Zombie-paper-Gambetta}
(and its several variations \cite{Xmon,Dicarlo}) has now become the
superconducting qubit of choice in many groups around the world due
to its simplicity of design and its superior coherence. Fidelities
of the single qubit gates are now routinely below \cite{Martinis-Nature,Sarah-Single-Qubit}
and those of the two-qubit gates are at the fault-tolerance threshold
levels required by the surface code \cite{Martinis-Nature,Sarah-CR,Maika-Parity}.
The challenge now is to scale the circuits up while maintaining and
improving further the qubit coherence times and gate fidelities \cite{Nick}.
Many important engineering problems however arise in the design of
larger multi-qubit devices such as signal crosstalk and qubit-qubit
crosstalk which show the need for better models/tools to understand
and improve the operation of the superconducting quantum processors. 

Several methods have been used to model and study the physics of superconducting
qubit circuits. The Jaynes-Cummings model \cite{Jaynes-Cummings}
originally introduced in quantum optics has routinely been applied
to the study of the so-called circuit-QED architecture \cite{Blais-xQED,Wallraff}
in which superconducting qubits are coupled to readout resonators
for their control and readout and two-qubit gate operations are mediated
by the bus resonators. Readout and bus resonators are typically designed
to be detuned away from the qubits to operate in the so-called dispersive
regime. In that regime one can eliminate the resonators up to desired
order in the bare qubit-resonator couplings and get an effective description
of the system in the qubit subspace. However calculation of the dispersive
quantities such as the exchange coupling or Purcell decay rates \cite{Controlling Spontaneous Emission - Houck}
of the qubits with the single mode Jaynes-Cummings model showed significant
discrepancy with the experimental measurements and attempts to include
higher harmonics of the resonators with multi-mode extensions of the
Jaynes-Cummings model failed due to divergence issues \cite{Bourassa-Multi-Mode-circuit-QED}.
\cite{Gely-Adrian-Solano} showed the convergence of the Lamb shift
in the specific case of a Josephson junction atom coupled to a multimode
resonator in the Rabi model. More recently \cite{Adrian-Long} studied
the convergence of the bare couplings between the superconducting
qubits and multimode resonators in various general coupling configurations.

Combination of lumped element circuit quantization methods \cite{Devoret-Les-Houches,BKD,Burkard}
with classical circuit synthesis techniques \cite{Foster,Brune,Newcomb}
resulted in ``blackbox quantization'' methods \cite{BBQ-Yale,Brune-Quantization,Solgun}
which allowed extraction of the parameters in the quantum Hamiltonian
models of the superconducting circuits from the electromagnetic finite-element
simulations. The simulations correspond to the linear passive part
of the circuits which is usually a distributed microwave structure
as seen looking into the ports defined across the Josephson junctions.
Although such an approach allows an accurate treatment of very general
structures consisting possibly of multiple microwave modes simulation
of large multi-qubit devices might quickly become computationally
demanding.

Following a similar approach we show here that for superconducting
processors consisting of low anharmocity qubits like transmons the
dispersive interaction parameters such as exchange coupling and Purcell
decay rates of the qubits and their coupling to the voltage drives
are related in a simple way to the microwave impedance response functions
as seen at the ``qubit ports'' and ``drive ports''. This reduces
a large portion of the design of multi-qubit superconducting devices
into a classical microwave engineering problem (up to the assumptions
and approximations we are making here) and allows one to avoid any
numerical multi-mode block-diagonalization or fitting of electromagnetic
finite-element simulations over a range of frequencies which are both
expensive if not prohibitive computational procedures.

We propose the following effective Hamiltonian to desribe a multi-qubit
superconducting device consisting of low anharmonicity qubits coupled
to each other and to the external world through a linear passive distributed
microwave structure:

\begin{widetext}

\begin{eqnarray}
\hat{\mathcal{H}}/\hbar & = & \stackrel[i=1]{N}{\sum}\omega_{i}\hat{b}_{i}^{\dagger}\hat{b}_{i}+\frac{\delta_{i}}{2}\hat{b}_{i}^{\dagger}\hat{b}_{i}(\hat{b}_{i}^{\dagger}\hat{b}_{i}-1)+\underset{i,j}{\sum}J_{ij}(\hat{b}_{i}\hat{b}_{j}^{\dagger}+\hat{b}_{i}^{\dagger}\hat{b}_{j})+\stackrel[i=1]{N}{\sum}\stackrel[d=1]{N_{D}}{\sum}\varepsilon_{id}(\hat{b}_{i}-\hat{b}_{i}^{\dagger})V_{d}\nonumber \\
 & + & \underset{i,k}{\sum}\chi_{ik}\hat{b}_{i}^{\dagger}\hat{b}_{i}\hat{a}_{k}^{\dagger}\hat{a}_{k}+\stackrel[k=1]{M}{\sum}\omega_{R_{k}}\hat{a}_{k}^{\dagger}\hat{a}_{k}+\frac{\chi_{kk}^{(R)}}{2}\hat{a}_{k}^{\dagger}\hat{a}_{k}(\hat{a}_{k}^{\dagger}\hat{a}_{k}-1)+\underset{k,k'}{\sum}J_{kk'}^{\left(R\right)}(\hat{a}_{k}\hat{a}_{k'}^{\dagger}+\hat{a}_{k}^{\dagger}\hat{a}_{k'})\label{eq:Hamiltonian}
\end{eqnarray}

\end{widetext}where we have $N$ qubit modes and $M$ resonator modes
represented as Duffing oscillators in the harmonic oscillator basis
and $N_{D}$ voltage drives. In the first line we have terms corresponding
to the qubit subspace: $\hat{b}_{i}^{(\dagger)}$ is the annihilation(creation)
operator of the qubit mode $i$ of frequency $\omega_{i}$ and anharmonicity
$\delta_{i}$. In the second line we have the resonator terms: $\hat{a}_{k}^{(\dagger)}$
is the annihilation(creation) operator of the resonator mode $k$
with frequency $\omega_{R_{k}}$ and anharmonicity (or self-Kerr)
$\chi_{kk}^{(R)}$ (We will be using the terms ``resonator'' and
``internal mode'' interchangibly below to refer to the microwave
modes of the distributed linear passive structure the qubits are connected
to). Such an approximate description in the harmonic basis is valid
for qubits with low anharmonicity $\delta_{i}\ll\omega_{i}$ such
as transmons. Qubit modes $i$ and $j$ are coupled to each other
at exchange coupling rate $J_{ij}$ and the only remaining interaction
between the qubit and resonator modes are the dispersive energy shifts
$\chi_{ik}$'s.

We show that the exchange coupling rate $J_{ij}$ between qubit modes
$i$ and $j$ in such an effective description is a simple function
of the impedance response defined between the \textcolor{black}{``qubit
ports''}

\begin{equation}
J_{ij}=-\frac{1}{4}\sqrt{\frac{\omega_{i}\omega_{j}}{L_{i}L_{j}}}\mathrm{Im}\left[\frac{Z_{ij}\left(\omega_{i}\right)}{\omega_{i}}+\frac{Z_{ij}\left(\omega_{j}\right)}{\omega_{j}}\right]\label{eq:The-impedance-formula-for-J}
\end{equation}
where $\omega_{i}$ is the frequency of the qubit $i$ given by $\omega_{i}=\omega_{J_{i}}-\frac{E_{C}^{(i)}/\hbar}{1-E_{C}^{(i)}/(\hbar\omega_{J_{i}})}$
with $\omega_{J_{i}}=1/\sqrt{L_{J_{i}}C_{i}}$ and $E_{C}^{(i)}=\frac{e^{2}}{2C_{i}}$
being the charging energy of the qubit $i$ of total shunt capacitance
$C_{i}$. $L_{i}$ and $L_{j}$ are the ``qubit inductances'' corresponding
to the qubits $i$ and $j$, respectively; related to the bare junction
inductances $L_{J_{i}}$'s by $L_{i}=L_{J_{i}}/(1-\frac{2E_{C}^{(i)}}{\hbar\omega_{i}}$)
such that $\omega_{i}=1/\sqrt{L_{i}C_{i}}$. $Z_{ij}\left(\omega\right)$
is the $\left(i,j\right)$-entry of the multiport impedance matrix
\textcolor{black}{$\mathbf{Z}(\omega)$ connecting $i^{th}$ qubit's
port to the $j^{th}$ qubit's port. Qubit ports are defined between
the terminals of the Josephson junctions; i.e. port voltages are voltages
developed across and the port currents are the currents flowing through
the Josephson junctions (See also Appendix \eqref{sub:Definition-of-the-ports-in-HFSS}
for how to define qubit ports as lumped ports in electromagnetic simulators).
The multiport impedance matrix $\mathbf{Z}(\omega)$ is to be computed
between the qubit ports with Josephson junctions removed. $\mathbf{Z}(\omega)$
then gives the response of the linear part of the circuit seen by
looking into the qubit ports; in particular $Z_{ij}(\omega)$ is the
voltage developed across $i^{th}$ qubit's port while a current of
unit magnitude and frequency $\omega$ is driving $j^{th}$ qubit's
port while all other qubit ports left open. We note here that the
formula in Eq. \eqref{eq:The-impedance-formula-for-J} holds in the
case of a distributed microwave structure consisting of multiple internal
modes(possibly infinite) coupling the qubits.}

\textcolor{black}{$V_{d}$ in Eq. \eqref{eq:Hamiltonian} is the voltage
source driving the $d$-th drive line for $1\leq d\leq N_{D}$ (Assuming
there are a total of $N_{D}$ lines driving the system as shown in
Fig. \ref{fig:Cauer-circuit-with-drives}) and $\varepsilon_{id}$
is the matrix entry giving the coupling of the qubit $i$ to the voltage
source $V_{d}$. In Section \eqref{sub:Derivation-of-the-classical-drive-cross-talk-appendix}
we show that (under the assumption that no off-chip crosstalk is happening
between the drive lines)}

\begin{equation}
\varepsilon_{id}=\sqrt{\frac{\omega_{i}}{2\hbar L_{i}}}\mathrm{Im}\left[Z_{i,p(d)}(\omega_{i})\right]\frac{e^{i\theta_{d}}C_{p(d)}}{\sqrt{1+\omega_{d}^{2}Z_{0}^{2}C_{p(d)}^{2}}}\label{eq:epsilon-matrix}
\end{equation}
where $\theta_{d}=\frac{\pi}{2}-\arctan(\omega_{d}Z_{0}C_{p(d)})$
and $Z_{i,p(d)}(\omega_{i})$ is the entry of the multiport impedance
matrix connecting the drive port(with port index $p(d)$) corresponding
to the voltage source $V_{d}$ (\textcolor{black}{for the definition
of drive ports see Section \eqref{sec:The-Classical-cross-talk} and
Appendix \eqref{sub:Definition-of-the-ports-in-HFSS})} to the qubit
port $i$ evaluated at the frequency $\omega_{i}$ of qubit $i$;
$\omega_{d}$ is the frequency of the voltage source $V_{d}$(assuming
a single tone sinusoidal signal), $Z_{0}$ is the characteristic impedance
of the drive lines which is typically $Z_{0}=50\Omega$ and $C_{p(d)}$
is the shunting capacitance of the drive port corresponding to the
voltage source $V_{d}$\textcolor{black}{. Since the drive ports are
defined where the drive lines reach the chip the factor $\mathrm{Im}\left[Z_{i,p(d)}(\omega_{i})\right]$
in Eq. \eqref{eq:epsilon-matrix} gives the classical crosstalk happening
at the trasition region where the lines land onto the chip or on the
chip. We also calculate below the following in units of $dB$ as a
measure of the classical crosstalk assuming similar values for qubit
parameters in Eq. \eqref{eq:epsilon-matrix}}

\textcolor{black}{
\begin{equation}
X_{ij}=20\mathrm{log}_{10}\left(\frac{\mathrm{Im}[Z_{i,d(j)}(\omega_{i})]}{\mathrm{Im}[Z_{j,d(j)}(\omega_{j})]}\right)
\end{equation}
where $d(j)$ is the port index of the drive of the qubit $j$ . $X_{ij}$
is the voltage crosstalk in $dB$ seen by qubit $i$ while driving
qubit $j$.}

\textcolor{black}{The resonance frequency $\omega_{R_{k}}$ of the
resonator $k$ gets the dispersive shift $\chi_{ik}$ depending on
the state of the qubit $i$. We calculate $\chi_{ik}$ in Section
\eqref{sec:Anharmonicities-Chi-shifts} similar to what has been done
in \cite{BBQ-Yale} by including the fourth order nonlinear terms
in the junction potentials}

\textcolor{black}{
\begin{equation}
\chi_{ik}=8\delta_{i}\left(\frac{g_{ik}\omega_{R_{k}}}{\omega_{R_{k}}^{2}-\omega_{i}^{2}}\right)^{2}
\end{equation}
where $\delta_{i}$ is the anharmonicity of the qubit mode $i$ given
in Eq. \eqref{eq:lambda-anharmonicity} as $\delta_{i}=-E_{C}^{(i)}(\omega_{J_{i}}/\omega_{i})^{2}$
and $g_{ik}$ is the bare coupling rate between the qubit mode $i$
and the resonator mode $k$ given in Eq. \eqref{eq:g-coupling} below.}

\textcolor{black}{We assume that the losses in the system are small;
in particular we neglect any internal loss. Hence $\mathrm{Im}[\mathbf{Z}(\omega)]$
describes the lossless part of the system to a very good approximation.
In Section \eqref{sec:Purcell-rates} we describe how to include the
effect of external losses due to the coupling to drive lines by computing
Purcell rates for the qubit modes. We show that the Purcell loss rate
$\frac{1}{T_{1}^{i,d}}$ of qubit $i$ due to the drive line $d$}

\begin{equation}
\frac{1}{T_{1}^{i,d}}=\frac{2}{L_{i}}\mathrm{Im}\left[Z_{i,p(d)}\left(\omega_{i}\right)\right]^{2}\frac{\omega_{i}^{2}Z_{0}C_{p(d)}^{2}}{1+\omega_{i}^{2}Z_{0}^{2}C_{p(d)}^{2}}
\end{equation}

We note here that all the dispersive rates of qubit-qubit interactions
and of interactions of qubits with the external electronics are functionals
of the the multiport impedance function $\mathbf{Z}(\omega)$ and
bare junction inductances $L_{J_{i}}$'s since the shunting capacitances
$C_{i}$'s of the qubit ports are related to the residue $\mathbf{A}_{0}$
of $\mathbf{Z}(\omega)$ at DC as given in Eq. \eqref{eq:A0-C0} (Same
argument applies to the shunt capacitances $C_{p(d)}$'s of the drive
ports) and the qubit frequencies $\omega_{i}$'s and anharmonicities
$\delta_{i}$'s are functions of qubit shunt capacitances and bare
junction inductances.

\textcolor{black}{$J_{kk'}^{\left(R\right)}$ in the second line in
Eq. \eqref{eq:Hamiltonian} are exchange coupling rates between resonator
modes mediated by the qubits. We note here that terms of the form
$\chi_{iikk'}\hat{b}_{i}^{\dagger}\hat{b}_{i}\hat{a}_{k}^{\dagger}\hat{a}_{k'}$
that are usually dropped by rotating wave approximation might be comparable
to other terms in Eq. \eqref{eq:Hamiltonian} if the frequencies $\omega_{R_{k}}$,
$\omega_{R_{k'}}$ of resonators $k$, $k'$ are not detuned enough.
In Eq. \eqref{eq:Hamiltonian} we also neglected drive terms on the
resonators.}

\section{\label{sec:Derivation-of-J-couplings}Derivation of the Formula for
the Exchange Coupling Rates between the Qubits}

Assuming we have $N$ Josephson junctions in the circuit we define
the $N\times N$ multiport impedance matrix $\mathbf{Z}$ seen looking
into qubit ports defined across the junction terminals ($\mathbf{Z}$
has to be evaluated without shunting the \textcolor{black}{qubit ports
by Josephson junctions}). Neglecting all the losses we can write the
following partial fraction expansion for the imaginary part of $\mathbf{Z}\left(\omega\right)$
as a function of the frequency variable $\omega$ \cite{Newcomb}

\begin{eqnarray}
\mathbf{Z}_{I}\left(\omega\right) & = & \mathrm{Im}\left[\mathbf{Z}\left(\omega\right)\right]\nonumber \\
 & = & -\frac{\mathbf{A}_{0}}{\omega}+\stackrel[k=1]{M}{\sum}\frac{\mathbf{A}_{k}\omega}{\omega_{R_{k}}^{2}-\omega^{2}}+\mathbf{A}_{\infty}\omega\label{eq:Partial-Fraction-Expansion}
\end{eqnarray}
where $\omega_{R_{k}}$'s are the frequencies of the internal modes
corresponding to readout and bus resonators and $\mathbf{A}_{k}$'s
are rank-1 \textcolor{black}{\cite{DDV}} real symmetric $N\times N$
matrices for $1\leq k\leq M$. Although we have truncated the part
corresponding to internal modes to $M$ terms as we will see below
the formula in Eq. \eqref{eq:The-impedance-formula-for-J} stays valid
in the limit of an infinite number of modes $M\rightarrow\infty$
(more generally one can think of the multiport impedance expansion
in Eq. \eqref{eq:Partial-Fraction-Expansion} as being corresponding
to any distributed electromagnetic structure seen by the junctions).

Starting with the expansion in Eq. \eqref{eq:Partial-Fraction-Expansion}
we can synthesize a lossless multiport lumped element circuit \cite{Newcomb}
\textcolor{black}{as shown in Fig. \eqref{fig:Cauer-circuit}. We
see $N$ qubit ports on the left in Fig. \eqref{fig:Cauer-circuit}
which are shunted by Josephson junctions. Using the method described
in \cite{Burkard} we can identify the degrees of freedom in this
circuit and derive the following Hamiltonian (see Appendix \eqref{sub:Derivation-of-the-Hamiltonian})}

\begin{equation}
\mathcal{H}=\frac{1}{2}\mathbf{Q}^{T}\mathbf{C}^{-1}\mathbf{Q}+\frac{1}{2}\mathbf{\Phi}^{T}\mathbf{M}_{0}\mathbf{\Phi}-\stackrel[i=1]{N}{\sum}E_{J_{i}}\cos\left(\varphi_{J_{i}}\right)\label{eq:initial-Hamiltonian}
\end{equation}
where $\mathbf{\Phi}=\left(\Phi_{J_{1}},\ldots,\Phi_{J_{N}},\Phi_{R_{1}},\ldots,\Phi_{R_{M}}\right)^{T}$
being the flux coordinate vector. $\varphi_{J_{i}}$ is the phase
of the junction $i$ related to the flux across it by the Josephson
relation $\Phi_{J_{i}}=\frac{\mathrm{\Phi}_{0}}{2\pi}\varphi_{J_{i}}$,
for $1\leq i\leq N$. $\Phi_{R_{k}}$ is the flux across the inductor
of the internal mode $k,$ $1\leq k\leq M$. $E_{J_{i}}$ is the Josephson
energy of junction $i$ related to its inductance $L_{J_{i}}$ by
$E_{J_{i}}=\left(\frac{\mathrm{\Phi}_{0}}{2\pi}\right)^{2}\frac{1}{L_{J_{i}}}$.
The capacitance matrix $\mathbf{C}$ is given by

\begin{equation}
\mathbf{C}=\left(\begin{array}{cc}
\mathbf{C}_{0} & -\mathbf{C}_{0}\mathbf{R}^{T}\\
-\mathbf{R}\mathbf{C}_{0} & \mathbf{1}_{M\times M}+\mathbf{R}\mathbf{C}_{0}\mathbf{R}^{T}
\end{array}\right)\label{eq:Capacitance-matrix}
\end{equation}
where $\mathbf{C}_{0}$ is diagonal with entries $\left(C_{1},\ldots,C_{N}\right)$,
$C_{i}$ being the total capacitance shunting the junction $i$. This
is a valid physical assumption since it corresponds to having no direct
electrostatic dipole-dipole interaction between junction terminals.
Such an assumption will keep our discussion simple although the case
of non-diagonal $\mathbf{C}_{0}$ will not change any of the results.
\textcolor{black}{In such a case one can treat the non-diagonal part
of $\mathbf{C}_{0}$ at frequency $\epsilon$ like the other terms
at finite frequencies $\omega_{R_{k}}$'s in the impedance expansion
in Eq. \eqref{eq:Partial-Fraction-Expansion} and apply the Scrieffer-Wolff
transformation as described below in the limit of $\epsilon\rightarrow0$
(A more rigorous algorithm in the case of non-diagonal $\mathbf{C}_{0}$
would be to remove as much diagonal part of $\mathbf{C}_{0}$ as possible
while keeping the rest still positive semidefinite and apply the small
$\epsilon$ frequency treatment we just described to an eigendecomposition
of the non-diagonal part). }

$\mathbf{R}$ is a $M\times N$ matrix generating the couplings between
qubits and internal modes. $\mathbf{R}$ consists of row vectors $r_{k}=\left(r_{k1}\ldots r_{kN}\right)$
with $r_{k}^{T}r_{k}=\mathbf{A}_{k}$. $\mathbf{M}_{0}$ matrix is
diagonal with entries $\left(1/L_{1},\ldots,1/L_{N},1/L_{R_{1}},\ldots,1/L_{R_{M}}\right)$
where $L_{R_{k}}=1/\omega_{R_{k}}^{2}$ for $1\leq k\leq M$. Here
we replaced the Josephson junction $i$ with the qubit inductance
$L_{i}$ such that $1/\sqrt{L_{i}C_{i}}=\omega_{i}$.\textcolor{red}{{}
}\textcolor{black}{An important point to note here is that the choice
of $L_{i}$ over the bare junction inductance $L_{J_{i}}$ makes the
two-body terms(that appear after expanding the nonlinear terms in
the junction potentials and normal ordering) in Eq. (10) of \cite{BBQ-Yale}
vanish up to the order of interest here. This is crucial since these
terms might contain significant residual couplings between qubit and
internal modes. We refer the reader to Appendix \eqref{sub:Expansion-of-the-junction-potentials}
for details.}

We do a capacitance rescaling \cite{Brito} $\mathbf{\Phi}_{J}\rightarrow\mathbf{C}_{0}^{1/2}\mathbf{\Phi}_{J}$
to transform the capacitance matrix $\mathbf{C}$ as follows

\begin{equation}
\mathbf{C}\rightarrow\left(\begin{array}{cc}
\mathbf{1}_{N\times N} & -\mathbf{C}_{0}^{1/2}\mathbf{R}^{T}\\
-\mathbf{R}\mathbf{C}_{0}^{1/2} & \mathbf{1}_{M\times M}+\mathbf{R}\mathbf{C}_{0}\mathbf{R}^{T}
\end{array}\right)
\end{equation}
and $\mathbf{M}_{0}$ transforming into the diagonal matrix with entries
$(\omega_{1}^{2},\ldots,\omega_{N}^{2},\omega_{R_{1}}^{2},\ldots,\omega_{R_{M}}^{2})$.
At this point \textcolor{black}{we note that the coupling $g_{ik}$
between the qubit mode $i$ and internal mode $k$ is given by}

\textcolor{black}{
\begin{equation}
g_{ik}=\frac{\sqrt{\omega_{i}\omega_{R_{k}}}}{2}r_{ki}\sqrt{C_{i}}\label{eq:g-coupling}
\end{equation}
where we also note that $r_{ki}\sqrt{C_{i}}$ is a small parameter
i.e. $r_{ki}\sqrt{C_{i}}\ll1$. }We then apply the transformation

\begin{equation}
\mathbf{T}=\left(\begin{array}{cc}
\mathbf{1}_{N\times N} & \mathbf{C}_{0}^{1/2}\mathbf{R}^{T}\\
\mathbf{0}_{M\times N} & \mathbf{1}_{M\times M}
\end{array}\right)\label{eq:T-transformation}
\end{equation}
to reduce the capacitance matrix to identity

\begin{equation}
\mathbf{C}\rightarrow\mathbf{T}^{T}\mathbf{C}\mathbf{T}=\mathbf{1}
\end{equation}

Then $\mathbf{M}_{0}$ transforms to $\mathbf{M}_{1}$ as

\begin{eqnarray}
\mathbf{M}_{1} & = & \mathbf{T}^{T}\mathbf{C}_{0}^{-1/2}\mathbf{M}_{0}\mathbf{C}_{0}^{-1/2}\mathbf{T}\nonumber \\
 & = & \left(\begin{array}{cc}
\mathbf{\Omega}_{J}^{2} & \mathbf{\Omega}_{J}^{2}\mathbf{C}_{0}^{1/2}\mathbf{R}^{T}\\
\mathbf{R}\mathbf{C}_{0}^{1/2}\mathbf{\Omega}_{J}^{2} & \mathbf{\Omega}_{R'}^{2}
\end{array}\right)\label{eq:M1-matrix}
\end{eqnarray}
where $\mathbf{\Omega}_{R'}^{2}=\mathbf{\Omega}_{R}^{2}+\mathbf{R}\mathbf{C}_{0}^{1/2}\mathbf{\Omega}_{J}^{2}\mathbf{C}_{0}^{1/2}\mathbf{R}^{T}$,
$\mathbf{\Omega}_{J}$ is diagonal with entries $\left(\omega_{1},\ldots,\omega_{N}\right)$,
$\omega_{i}=1/\sqrt{L_{i}C_{i}}$ for $1\leq i\leq N$ and $\mathbf{\Omega}_{R}$
is diagonal with entries $\left(\omega_{R_{1}},\ldots,\omega_{R_{M}}\right)$.
\textcolor{black}{Here we observe that the resonator frequencies get
small corrections that we will neglect in the following and the couplings
in between the modes in the resonator subspace are of order $\Omega_{J}(g/\Omega_{R})^{2}$
where $g$ is the bare coupling strength between qubit and resonator
modes. The resonator subspace being diagonal to order $\Omega_{J}(g/\Omega_{R})^{2}$
is important in the application of the Schrieffer-Wolff transformation
below as it allows to capture small couplings by only a second order
Schrieffer-Wolff transformation that would otherwise require higher
order corrections.}

We now block-diagonalize $\mathbf{M}_{1}$ by applying a Schrieffer-Wolff
transformation to get

\begin{equation}
\widetilde{\mathbf{M}}_{1}=\exp\left(-\mathbf{S}\right)\mathbf{M}_{1}\exp\left(\mathbf{S}\right)\label{eq:block-diagonal-M1}
\end{equation}
where $\mathbf{S}$ is skew-symmetric and $\widetilde{\mathbf{M}}_{1}$
block-diagonal which can be computed up to desired order in the bare
couplings using Eqs. (B.12) and (B.15) in \cite{Winkler}. We note
that since this transformation is unitary it will keep the capacitance
matrix identity such that we have the following block-diagonal Hamiltonian
in the final frame

\begin{equation}
\mathcal{H}=\frac{1}{2}\boldsymbol{q}^{T}\boldsymbol{q}+\frac{1}{2}\boldsymbol{\phi}^{T}\widetilde{\mathbf{M}}_{1}\boldsymbol{\phi}+\mathcal{O}(\varphi_{J}^{4})\label{eq:block-diagonal-hamiltonian}
\end{equation}
where the final coordinate fluxes $\boldsymbol{\phi}$ are related
to the initial coordinates $\mathbf{\Phi}$ by

\begin{equation}
\mathbf{\Phi}=\left(\begin{array}{c}
\mathbf{\Phi}_{J}\\
\mathbf{\Phi}_{R}
\end{array}\right)=\left(\begin{array}{cc}
\mathbf{C}_{0}^{-1/2} & \mathbf{0}\\
\mathbf{0} & \mathbf{1}
\end{array}\right)\mathbf{T}\exp(\mathbf{S})\boldsymbol{\phi}
\end{equation}
and $\mathcal{O}(\varphi_{J}^{4})$ term standing for higher order
nonlinear corrections giving anharmonicities and dispersive shifts
between modes calculated in Appendix \eqref{sub:Expansion-of-the-junction-potentials}.

Using Eq. (B.15c) in \cite{Winkler}, to second order in the bare
couplings

\begin{equation}
\left(\widetilde{\mathbf{M}}_{1}\right)_{ij}=\frac{1}{2}\underset{k}{\sum}\left(\mathbf{M}_{1}\right)_{ik}\left(\mathbf{M}_{1}\right)_{kj}\left[\frac{1}{\omega_{i}^{2}-\omega_{R_{k}}^{2}}+\frac{1}{\omega_{j}^{2}-\omega_{R_{k}}^{2}}\right]\label{eq:M1-tilde-ij}
\end{equation}
where again $i$ and $j$ are qubit labels and $k$ labels internal
modes. $\left(\mathbf{M}_{1}\right)_{ik}$ is the $\left(i,k\right)$-th
entry of the matrix $\mathbf{M}_{1}$ and from Eq. \eqref{eq:M1-matrix}
we have

\begin{equation}
\left(\mathbf{M}_{1}\right)_{ik}=\omega_{i}^{2}C_{i}^{1/2}r_{ki}\label{eq:Mik}
\end{equation}
Noting again $r_{k}^{T}r_{k}=\mathbf{A}_{k}$ we can write

\begin{eqnarray}
\left(\mathbf{M}_{1}\right)_{ik}\left(\mathbf{M}_{1}\right)_{kj} & = & \omega_{i}^{2}\omega_{j}^{2}C_{i}^{1/2}C_{j}^{1/2}r_{ki}r_{kj}\nonumber \\
 & = & \omega_{i}^{2}\omega_{j}^{2}C_{i}^{1/2}C_{j}^{1/2}\left(\mathbf{A}_{k}\right)_{ij}
\end{eqnarray}
Hence we can re-write Eq. \eqref{eq:M1-tilde-ij}

\begin{eqnarray}
\left(\widetilde{\mathbf{M}}_{1}\right)_{ij} & = & \frac{1}{2}\underset{k}{\sum}\omega_{i}^{2}\omega_{j}^{2}C_{i}^{1/2}C_{j}^{1/2}\left[\frac{\left(\mathbf{A}_{k}\right)_{ij}}{\omega_{i}^{2}-\omega_{R_{k}}^{2}}+\frac{\left(\mathbf{A}_{k}\right)_{ij}}{\omega_{j}^{2}-\omega_{R_{k}}^{2}}\right]\nonumber \\
 & = & -\frac{1}{2}\omega_{i}^{2}\omega_{j}^{2}C_{i}^{1/2}C_{j}^{1/2}\mathrm{Im}\left[\frac{Z_{ij}(\omega_{i})}{\omega_{i}}+\frac{Z_{ij}(\omega_{j})}{\omega_{j}}\right]
\end{eqnarray}

Quantizing the system by introducing the annihilation and creation
operators for the qubit modes in the final frame by $\hat{\phi}_{i}=\sqrt{\frac{\hbar Z_{i}}{2}}(\hat{b}_{i}+\hat{b}_{i}^{\dagger})$
for $1\leq i\leq N$ and noting that the characteristic impedance
$Z_{i}$ for the qubit mode $i$ is $Z_{i}=1/\omega_{i}$ in that
frame

\begin{eqnarray}
J_{ij} & = & \frac{1}{2}\sqrt{Z_{i}Z_{j}}\left(\widetilde{\mathbf{M}}_{1}\right)_{ij}\nonumber \\
 & = & -\frac{1}{4}\sqrt{\frac{\omega_{i}\omega_{j}}{L_{i}L_{j}}}\mathrm{Im}\left[\frac{Z_{ij}\left(\omega_{i}\right)}{\omega_{i}}+\frac{Z_{ij}\left(\omega_{j}\right)}{\omega_{j}}\right]
\end{eqnarray}
$J_{ij}$ in the above formula is in the units of radians per second.
We note that this formula takes into account all the modes(possibly
infinite) that might be present in the electromagnetic structure coupling
the qubits.

\subsection{Example 1: Two transmons coupled through a single mode $LC$ resonator
bus}

\begin{figure}
\begin{centering}
\includegraphics[scale=0.6]{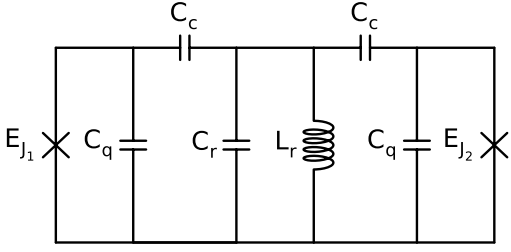}
\par\end{centering}

\caption{\label{fig:2-Q-1-bus}Example circuit of two transmons capacitively
coupled through a single mode bus. Both transmons have the same shunting
capacitance $C_{q}$ and the same coupling capacitances $C_{c}$ to
the bus.}
\end{figure}

In this section we will apply the formula in Eq. $\eqref{eq:The-impedance-formula-for-J}$
for the $J$-coupling rate derived in the previous section to the
simple circuit of two transmons coupled through a lumped $LC$ resonator
as shown in Fig. \eqref{fig:2-Q-1-bus} and compare it to the expression
derived in \cite{Jay-Juelich}: 

\begin{equation}
J=\frac{g_{1}g_{2}\left(\omega_{1}+\omega_{2}-2\omega_{r}\right)}{2\left(\omega_{1}-\omega_{r}\right)\left(\omega_{2}-\omega_{r}\right)}\label{eq:Jpert}
\end{equation}
where $g_{1}$, $g_{2}$ are couplings of qubits $1,2$ to the bus,
$\omega_{1}$, $\omega_{2}$ and $\omega_{r}$ are qubit and resonator
frequencies; respectively.

\begin{figure}
\begin{centering}
\includegraphics[scale=0.2]{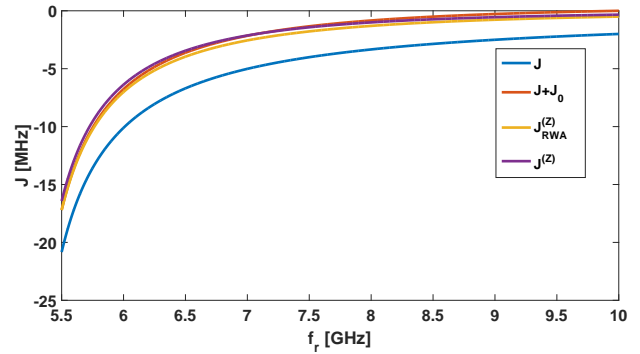}
\par\end{centering}

\caption{\label{fig:compare-formulas-plot}Comparison of $J$-coupling expressions
$J^{\left(Z\right)},$ $J_{RWA}^{\left(Z\right)}$, $J$ and $J$+$J_{0}$
for bus frequency $f_{r}$ ranging from $5.5\,GHz$ to $10\,GHz$
for the circuit in Fig. \eqref{fig:2-Q-1-bus} with the following
set of parameter values $g_{1}=g_{2}=100\,MHz$, $\omega_{1}=2\pi(4.90\,GHz)$
and $\omega_{2}=2\pi\left(5.10\,GHz\right)$, $\delta_{1}=\delta_{2}=-340\,MHz$.
Vertical axis is $J$-coupling rate in $MHz$.}
\end{figure}

The circuit in Fig. \eqref{fig:2-Q-1-bus} has the following Hamiltonian

\begin{equation}
H=\frac{1}{2}\mathbf{Q}^{T}\mathbf{C}^{-1}\mathbf{Q}+\frac{1}{2}\mathbf{\Phi}^{T}\mathbf{M}_{0}\mathbf{\Phi}-E_{J_{1}}\cos\left(\varphi_{1}\right)-E_{J_{2}}\cos\left(\varphi_{2}\right)
\end{equation}
where

\begin{eqnarray}
\mathbf{C} & = & \left(\begin{array}{ccc}
C_{q}+C_{c} & 0 & -C_{c}\\
0 & C_{q}+C_{c} & -C_{c}\\
-C_{c} & -C_{c} & C_{r}+2C_{c}
\end{array}\right)\label{eq:C-2-Q-1-B}
\end{eqnarray}
$\mathbf{M}_{0}$ diagonal with entries $\left(0,0,1/L_{r}\right)$
and the coordinate vector $\mathbf{\Phi}=(\Phi_{J_{1}},\Phi_{J_{2}},\Phi_{r})^{T}$
holds the fluxes across the inductive branches. Typically $C_{c}\ll C_{q}\ll C_{r}$
holds so that we can approximately write

\begin{equation}
\mathbf{C}^{-1}\cong\left(\begin{array}{ccc}
1/C_{q} & \frac{C_{c}^{2}}{C_{q}^{2}C_{r}} & \frac{C_{c}}{C_{q}C_{r}}\\
\frac{C_{c}^{2}}{C_{q}^{2}C_{r}} & 1/C_{q} & \frac{C_{c}}{C_{q}C_{r}}\\
\frac{C_{c}}{C_{q}C_{r}} & \frac{C_{c}}{C_{q}C_{r}} & 1/C_{r}
\end{array}\right)\label{eq:Cinv-2-Q-1-B}
\end{equation}
so that we have

\begin{eqnarray}
g_{1} & = & \frac{1}{2\sqrt{Z_{1}Z_{r}}}\frac{C_{c}}{C_{q}C_{r}}\\
g_{2} & = & \frac{1}{2\sqrt{Z_{2}Z_{r}}}\frac{C_{c}}{C_{q}C_{r}}
\end{eqnarray}
where $Z_{i}=\sqrt{L_{i}/C_{q}}$ and $Z_{r}=\sqrt{L_{r}/C_{r}}$.
We note here that although there is no direct electrostatic dipole
coupling between qubits in Eq. \eqref{eq:C-2-Q-1-B} a mediated coupling
$J_{0}$ appears in Eq. \eqref{eq:Cinv-2-Q-1-B}. As we will see below
the magnitude of $J_{0}$ is non-negligible compared to $J$ in Eq.
\eqref{eq:Jpert} hence one should compute $J+J_{0}$ for the total
exchange coupling rate as we did in Fig. \eqref{fig:compare-formulas-plot}.
We note that

\begin{eqnarray}
J_{0} & = & \frac{1}{2\sqrt{Z_{1}Z_{2}}}\frac{C_{c}^{2}}{C_{q}^{2}C_{r}}\nonumber \\
 & = & \frac{2}{\omega_{r}}g_{1}g_{2}\label{eq:J0}
\end{eqnarray}

We now apply the impedance formula for the $J$-coupling in Eq. \eqref{eq:The-impedance-formula-for-J}
to the circuit in Fig. \eqref{fig:2-Q-1-bus}. We need to first compute
the two-port impedance matrix between the ports shunted by Josephson
junctions. This can be done by an $ABCD$-matrix analysis \cite{Pozar},
for example. One then gets

\begin{equation}
\mathrm{Im}\left[Z_{12}\left(\omega\right)\right]=\frac{C_{c}^{2}L_{r}\omega/\left(C_{q}+C_{c}\right)}{C_{q}\left(1-\omega^{2}/\omega_{r}^{2}\right)+C_{c}\left(1-2\omega^{2}/\omega_{qr}^{2}-\omega^{2}/\omega_{r}^{2}\right)}
\end{equation}
where $\omega_{r}=1/\sqrt{L_{r}C_{r}}$ and $\omega_{qr}=1/\sqrt{L_{r}C_{q}}$.
We note that in actual devices $C_{q}\ll C_{r}$ hence $\omega_{r}\ll\omega_{qr}$.
We can then neglect the term $-2\omega^{2}/\omega_{qr}^{2}$ appearing
in the denominator compared to the term $-\omega^{2}/\omega_{r}^{2}$
such that

\begin{equation}
\mathrm{Im}\left[Z_{12}\left(\omega\right)\right]\cong\frac{C_{c}^{2}L_{r}\omega}{\left(C_{q}+C_{c}\right)^{2}\left(1-\omega^{2}/\omega_{r}^{2}\right)}
\end{equation}
Noting also $C_{c}\ll C_{q}$ we have

\begin{eqnarray}
\mathrm{Im}\left[Z_{12}\left(\omega\right)\right] & \cong & \frac{C_{c}^{2}L_{r}\omega}{C_{q}^{2}\left(1-\omega^{2}/\omega_{r}^{2}\right)}\nonumber \\
 & = & \frac{1}{2}\frac{C_{c}^{2}L_{r}\omega_{r}\omega}{C_{q}^{2}}\left(\frac{1}{\omega_{r}-\omega}+\frac{1}{\omega_{r}+\omega}\right)\label{eq:ImZ12-exp}
\end{eqnarray}
hence by Eq. \eqref{eq:The-impedance-formula-for-J}

\begin{align}
J^{\left(Z\right)} & =-\frac{1}{8}\frac{\sqrt{\omega_{1}\omega_{2}}}{\sqrt{L_{1}L_{2}}}\frac{C_{c}^{2}L_{r}\omega_{r}}{C_{q}^{2}}\left(\frac{1}{\omega_{r}-\omega_{1}}+\frac{1}{\omega_{r}-\omega_{2}}+\right.\nonumber \\
 & +\left.\frac{1}{\omega_{r}+\omega_{1}}+\frac{1}{\omega_{r}+\omega_{2}}\right)\label{eq:JZ}
\end{align}
where we used the superscript $Z$ to indicate the application of
the impedance $J$-coupling formula in Eq. \eqref{eq:The-impedance-formula-for-J}.

If we interpret the first two terms inside the paranthesis in Eq.
\eqref{eq:JZ} as the RWA-terms we can write

\begin{eqnarray}
J_{RWA}^{\left(Z\right)} & =- & \frac{1}{8}\frac{\sqrt{\omega_{1}\omega_{2}}}{\sqrt{L_{1}L_{2}}}\frac{C_{c}^{2}L_{r}\omega_{r}}{C_{q}^{2}}\left(\frac{1}{\omega_{r}-\omega_{1}}+\frac{1}{\omega_{r}-\omega_{2}}\right)\nonumber \\
 & = & \left(\frac{\omega_{1}\omega_{2}}{\omega_{r}^{2}}\right)\frac{g_{1}g_{2}\left(\omega_{1}+\omega_{2}-2\omega_{r}\right)}{2\left(\omega_{1}-\omega_{r}\right)\left(\omega_{2}-\omega_{r}\right)}\nonumber \\
 & = & \left(\frac{\omega_{1}\omega_{2}}{\omega_{r}^{2}}\right)J\label{eq:JZ-RWA}
\end{eqnarray}

We note here that the standard expression for the exchange coupling
$J$ in Eq. \eqref{eq:Jpert} is obtained with a RWA; this is why
we only kept the first two terms inside the paranthesis in Eq. \eqref{eq:JZ}
and defined $J_{RWA}^{\left(Z\right)}$ in Eq. \eqref{eq:JZ-RWA}.

We now compare the formulas obtained above in Fig. \eqref{fig:compare-formulas-plot}
with the following set of realistic parameter values $g_{1}=g_{2}=100\,MHz$,
$\omega_{1}=2\pi(4.90\,GHz)$ and $\omega_{2}=2\pi\left(5.10\,GHz\right)$,
$\delta_{1}=\delta_{2}=-340\,MHz$.

\subsection{Example 2: Scaling of $J$ coupling rates in a multi-qubit device}

\begin{figure}
\begin{centering}
\includegraphics[scale=0.35]{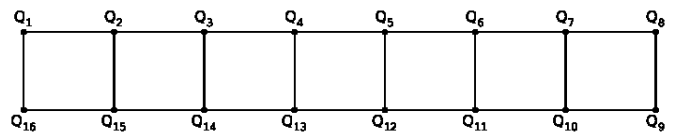}
\par\end{centering}

\caption{\label{fig:2x8-device}2x8 device connectivity: 16 qubits are arranged
in 2 rows. Nodes respresent the qubits while edges linking the qubits
represent buses. There are two qubits connected to each bus and there
is a total of $22$ buses.}
\end{figure}

\begin{figure}
\begin{centering}
\includegraphics[scale=0.45]{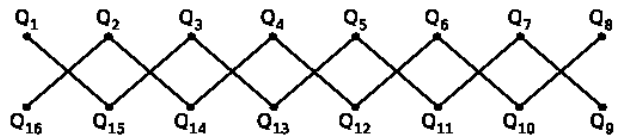}
\par\end{centering}

\caption{\label{fig:2x8-device-4QbyBus}2x8 device with four qubits per bus
arrangement. Scaling of the $J$ couplings over the lattice is compared
to the arrangement in Fig. \eqref{fig:2x8-device}. Crossed links
represent bus resonators each connected to 4 qubits and there are
$7$ buses in total.}
\end{figure}

\begin{figure}
\begin{centering}
\includegraphics[scale=0.2]{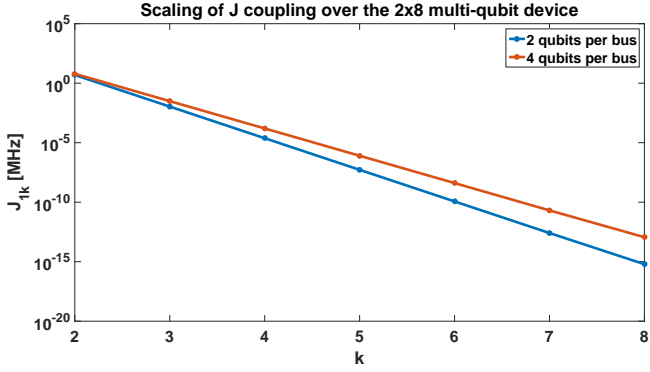}
\par\end{centering}

\centering{}\caption{\label{fig:2by8-J-scaling-plot}Exponential decay of the $J_{1k}$
coupling rate for $k=2,\ldots8$ as measured from the first qubit
$Q_{1}$ to the right in the upper rows in Figs. \eqref{fig:2x8-device}
and \eqref{fig:2x8-device-4QbyBus} as a function of qubit index $k$.
$J_{12}$ is $-4.9\,MHz$ and $-6.1\,MHz$ in the devices in Figs.
\eqref{fig:2x8-device} and \eqref{fig:2x8-device-4QbyBus}, respectively.}
\end{figure}

In this section we apply the impedance formula in Eq. \eqref{eq:The-impedance-formula-for-J}
for the exchange couplings $J_{ij}$ to the multi-qubit device shown
in Fig. \eqref{fig:2x8-device} to calculate the decay of $J$ over
the chip. This is a simplified model of an actual multi-qubit device
recently released by IBM in its online cloud environment for quantum
computing: IBM Q Experience \cite{IBM-Q-Experience}. The device consists
of $16$ qubits arranged in two rows and connected to each other by
$22$ bus resonators with two qubits per bus. To compare we also apply
the impedance formula for $J$ coupling to the arrangement shown in
Fig. \eqref{fig:2x8-device-4QbyBus} where we have four qubits on
each bus. We model each bus as a simple $LC$ resonator at $6.30\,GHz$
capacitively coupled to qubits. Using realistic parameter values corresponding
to a real device fabricated at IBM we obtain the decay plots in Fig.
\eqref{fig:2by8-J-scaling-plot} which confirm exponential decay of
$J$ couplings over the chips.

\section{\label{sec:Voltage-couplings}Couplings of the Qubits to the Voltage
Drives}

Qubits are coupled to room temperature electronics for their readout
and control. Readout and control signals pass through several amplification/attenuation
stages as they travel through different stages in a dilution fridge.
In between these stages they are carried over transmission lines like
coaxial cables or the lines on a printed circuit board. We will content
ourselves here with modeling this coupling mechanism simply by voltage
sources driving the quantum chip through transmission lines(which
we assume to be inifinite in extent to keep things simple here and
represent them simply by resistors $Z_{0}$'s) as shown in Fig. \eqref{fig:Cauer-circuit-with-drives}.
This circuit is an augmented version of the multiport canonical circuit
in Fig. \eqref{fig:Cauer-circuit} where $N_{D}$ ``drive ports''
are added. The drive ports are defined at positions where drive lines
reach the chip (see Appendix \eqref{sub:Definition-of-the-ports-in-HFSS}
for more details on how to define the drive ports in an 3D finite-element
electromagnetic simulator). They are connected to transmission lines
of characteristic impedance $Z_{0}$ (typically $Z_{0}=50\Omega$)
which in turn are shunted by the voltage sources $V_{d}$ for $1\leq d\leq N_{D}$.
Such a simple circuit model will allow us to derive expressions for
the couplings $\varepsilon_{id}$ of the qubits to voltage drives
in this section. A similar analysis in Section \eqref{sec:Purcell-rates}
will allow us to compute Purcell loss rates of the qubit modes due
to their coupling to the drive lines.

As we show in Appendix \eqref{sub:Derivation-of-the-classical-drive-cross-talk-appendix}
the circuit in Fig. \eqref{fig:Cauer-circuit-with-drives} has the
following Hamiltonian given in Eq. \eqref{eq:Hamiltonian-drive-final-frame}
in the final block-diagonalized frame corresponding to $\widetilde{\mathbf{M}}_{1}$
in Eq. \eqref{eq:block-diagonal-hamiltonian}

\begin{equation}
H=\frac{1}{2}\left(\boldsymbol{q}-\mathbf{C}_{q}*\mathbf{V}_{V}\right)^{T}\left(\boldsymbol{q}-\mathbf{C}_{q}*\mathbf{V}_{V}\right)+\frac{1}{2}\boldsymbol{\phi}^{T}\widetilde{\mathbf{M}}_{1}\boldsymbol{\phi}+\mathcal{O}(\boldsymbol{\varphi}_{J}^{4})
\end{equation}
where the $(N+M)\times N_{D}$ matrix $\mathbf{C}_{q}$ gives the
coupling of the voltage sources $\mathbf{V}_{V}=(V_{1},\ldots,V_{N_{D}})$
to the charge degrees of freedom $\boldsymbol{q}$ of the circuit.
After quantizing this Hamiltonian by introducing the harmonic mode
operators $\hat{q}_{i}=-i\sqrt{\frac{\hbar}{2Z_{i}}}(\hat{b}_{i}-\hat{b}_{i}^{\dagger})$
for the qubit modes and computing the projection of $\mathbf{C}_{q}$
onto the qubit subspace one obtains the following drive term acting
in the qubit subspace

\begin{eqnarray}
H_{id}^{D} & = & i\sqrt{\frac{\hbar\omega_{i}}{2L_{i}}}\mathrm{Im}\left[Z_{i,p(d)}(\omega_{i})\right]\frac{C_{p(d)}V_{d}(\hat{b}_{i}-\hat{b}_{i}^{\dagger})}{1+i\omega_{d}Z_{0}C_{p(d)}}\label{eq:HD-id}
\end{eqnarray}
from which we get

\begin{equation}
\varepsilon_{id}=\sqrt{\frac{\omega_{i}}{2\hbar L_{i}}}\mathrm{Im}\left[Z_{i,p(d)}(\omega_{i})\right]\frac{e^{i\theta_{d}}C_{p(d)}}{\sqrt{1+\omega_{d}^{2}Z_{0}^{2}C_{p(d)}^{2}}}\label{eq:eps-ij}
\end{equation}
for the coupling matrix $\varepsilon_{id}$ appearing in the Hamiltonian
in Eq. \eqref{eq:Hamiltonian} and giving the coupling of the qubit
modes to the voltage drives. Here $\theta_{d}=\frac{\pi}{2}-\arctan(\omega_{d}Z_{0}C_{p(d)})$
and $Z_{i,p(d)}$ is the impedance entry connecting the qubit port
$i$ to drive port(with port index $p(d)$) corresponding to the voltage
source $V_{d}$. $C_{p(d)}$ is the total capacitance shunting the
$d$-th drive port, $\omega_{d}$ is the frequency of the signal driving
the qubit $j$ and $Z_{0}$ is the characteristic impedance of the
drive lines (typically $Z_{0}=50\Omega$). The last factor in Eq.
\eqref{eq:HD-id} is just a voltage division factor giving how much
of the drive voltage $V_{d}$ is seen across the $d$-th drive port.
The factor $\mathrm{Im}\left[Z_{i,p(d)}(\omega_{i})\right]$ gives
on the other hand the classical crosstalk.

\subsection{\label{sec:The-Classical-cross-talk}The classical crosstalk and
the location of the drive ports}

We define the classical crosstalk as the unwanted drive the qubit
$i$ experiences when we excite the device only \textcolor{black}{through
}the drive line of the qubit $j$. For the purpose of understanding
the classical cross-talk we will be only interested in the relative
magnitudes of the voltages seen by different qubits and according
to the analysis in Appendix \eqref{sub:Derivation-of-the-classical-drive-cross-talk-appendix}

\begin{equation}
X_{ij}=20\mathrm{log}_{10}\left(\frac{\mathrm{Im}[Z_{i,d(j)}(\omega_{i})]}{\mathrm{Im}[Z_{j,d(j)}(\omega_{j})]}\right)\label{eq:Classical-crosstalk-formula}
\end{equation}
is a good measure of the classical cross-talk in units of $dB$. Here
$Z_{i,d(j)}(\omega_{i})$ is the impedance entry connecting the drive
port $d(j)$ of the qubit $j$ to the qubit port $i$.

Although we have already stated in the previous sections that we defined
the drive ports where the drive lines reach the chip we give a more
precise description here on how we choose the locations of the drive
ports. As the drive signals travel over the transmission lines towards
the chip they will eventually reach the transition region (before
launching onto the chip) where they will no longer see a constant
impedance but a discontinuity off which some portion of the signal
will be reflected back. Ideally one would like to define the drive
ports at positions where this discontinuity first starts to appear.
The exact positions can be determined with a TDR (time-domain reflectometry)
measurement/simulation for example. In the absence of TDR information
one can make a safe choice by keeping the drive ports far enough from
the chip boundary. In electromagnetic finite-element simulators such
ports will typically be defined as wave ports on the planes (perpendicular
to the direction of propagation) in the cross-sections of the drive
lines. Such a choice for the drive ports will include any crosstalk
happening in the transition region (such as a spurious chip boundary
mode \cite{Wirebond-Crosstalk-Martinis} for example) in our crosstalk
measures defined above. See Appendix \eqref{sub:Definition-of-the-ports-in-HFSS}
for more details on how to define the drive ports in electromagnetic
finite-element simulators.

\subsection{Example: Classical crosstalk in a multi-qubit device}

\begin{figure}
\begin{raggedright}
\includegraphics[scale=0.35]{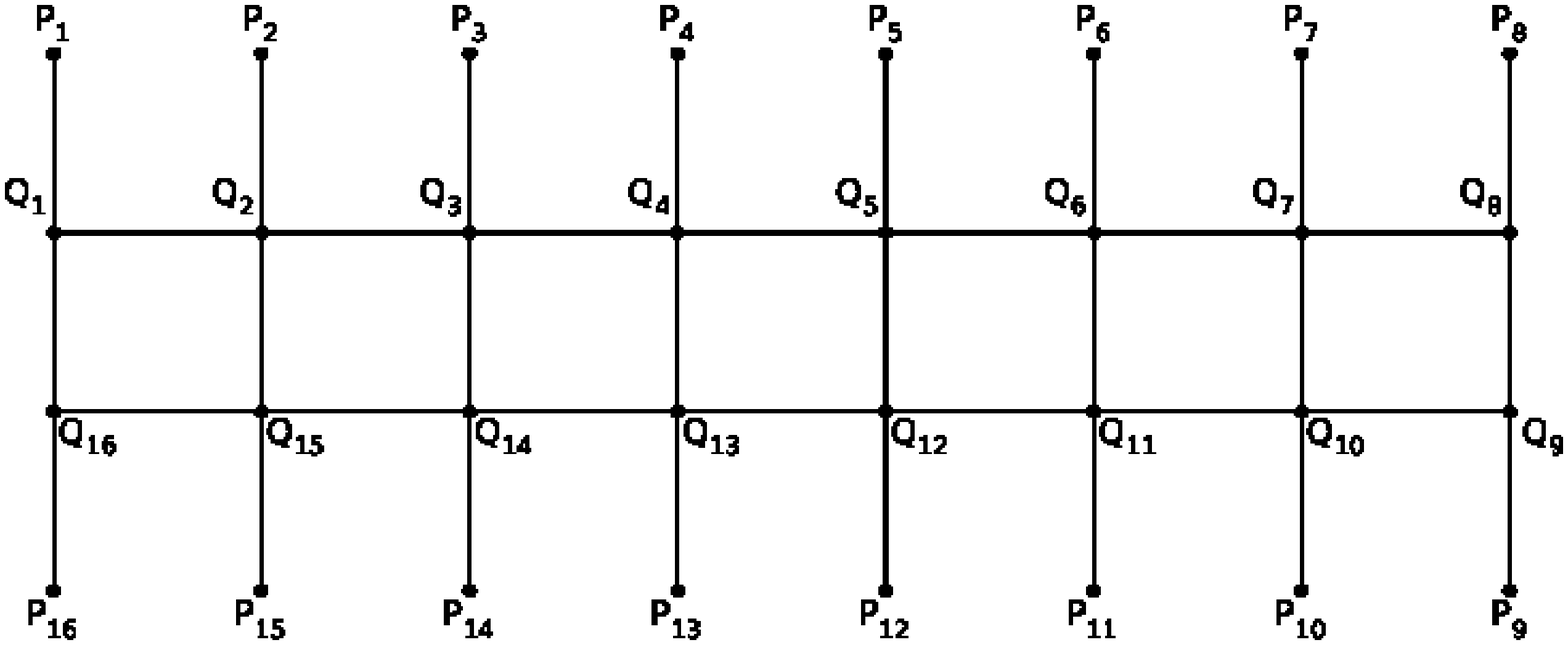}\\
\includegraphics[scale=0.2]{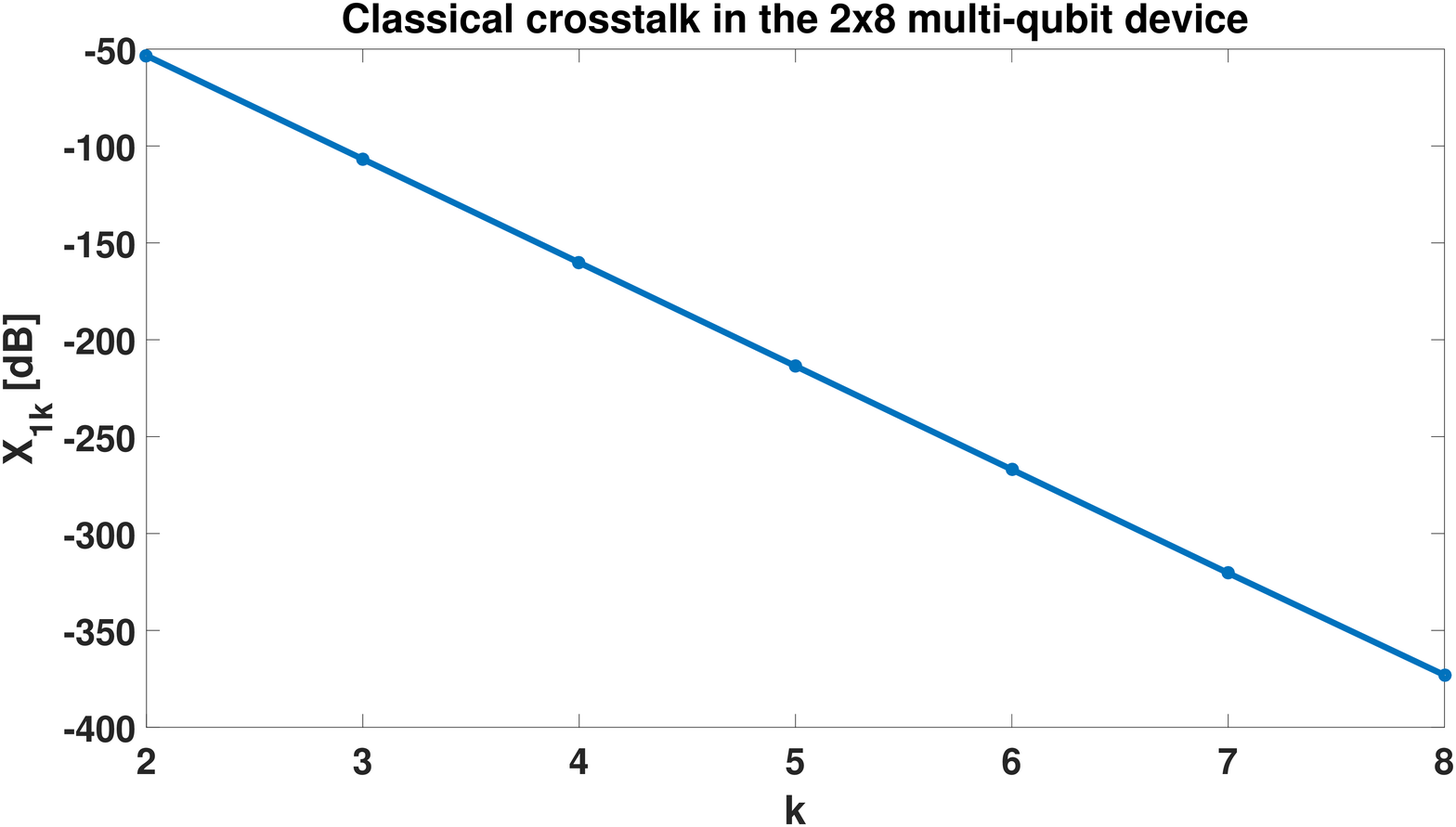}
\par\end{raggedright}

\caption{\label{fig:2x8-device-model-with-drives}Top: Augmented 2x8 device
model with readout resonators and drive ports added. Readout resonators
are represented with edges linking qubits $Q_{k}$'s to their drive
ports $P_{k}$'s. Bottom: How much of the drive voltage leaks into
other qubits in the upper row in the 2x8 multi-qubit device on the
top figure above when only the qubit $Q_{1}$ is excited through its
drive port $P_{1}$: classical crosstalk decays exponentially as a
function of distance from the qubit $Q_{1}$ to the right in the first
row of the circuit in the figure on the top. }
\end{figure}

In this section we augment our model for the 2x8 multi-qubit device
by adding the readout resonators and the drive ports as shown on the
left in Fig. \eqref{fig:2x8-device-model-with-drives} and apply the
the formula in Eq. \eqref{eq:Classical-crosstalk-formula} to evaluate
the cross-talk in the device. We plot $X_{1k}$ which gives the crosstalk
between the drive line of the qubit $Q_{1}$ and the other qubits
on the first row in Fig. \eqref{fig:2x8-device-model-with-drives}
as a function of the qubit label $k=2,\ldots,8$ in Fig. \eqref{fig:2x8-device-model-with-drives}.

\section{\label{sec:Purcell-rates}Purcell Loss Rates of The Qubit Modes}

Qubits are coupled to external electronics for their readout and control.
In Section \ref{sec:Voltage-couplings} we analyzed couplings of the
qubits to voltage drives. The same coupling mechanism causes relaxation
of the excitations in the qubit modes which is called the ``Purcell
Loss''. In this section we compute rates for the Purcell loss of
the qubit modes we identified in the earlier sections. 

As in Section \eqref{sec:Voltage-couplings} the coupling of the qubits
to external electronics is modeled with the idealized circuit model
in Fig. \eqref{fig:Cauer-circuit-with-drives} and we will use the
same coupling matrices of the formalism in \cite{Burkard} that we
calculated in Appendix \eqref{sub:Derivation-of-the-classical-drive-cross-talk-appendix}
for the drive couplings. We have $N_{D}$ baths corresponding to tranmission
lines of characteristic impedances $Z_{0}$'s driving the qubits as
shown in Fig. \eqref{fig:Cauer-circuit-with-drives}. Assuming couplings
of qubits to the lines are small, to first order in these couplings,
we will assume that $T_{1}$ rates can be computed separately for
each bath. The total rate will then be the sum of rates due to each
line.

We start by noting that when we have only the bath due to the drive
line of the voltage source $V_{d}$ with port index $p(d)$ $\mathbf{C}_{D}$
defined in Eq. \eqref{eq:Ck-def} is a scalar $C_{p(d)}$ for $1\leq d\leq N_{D}$.
Hence $\bar{\mathbf{m}}$ in Eq. \eqref{eq:mbar-before-Schr-Wollf}
is

\begin{equation}
\bar{\mathbf{m}}_{d}=-C_{p(d)}\left(\begin{array}{c}
\mathbf{0}_{N\times1}\\
\mathbf{v}_{d}
\end{array}\right)
\end{equation}
where $\mathbf{v}_{d}=(\begin{array}{ccc}
v_{1d} & \ldots & v_{Md}\end{array})^{T}$ is the $d$-th column of the matrix $\mathbf{V}$ corresponding to
the drive line with port index $p(d)$. After the Schrieffer-Wolff
transformation by Eq. \eqref{eq:D_ij-equation}

\begin{equation}
m_{id}=C_{p(d)}\mathrm{Im}\left[Z_{i,p(d)}\left(\omega_{i}\right)\right]/\sqrt{L_{i}}\label{eq:mij-final-frame}
\end{equation}
where $m_{id}$ is the coupling of the bath due to the $d$-th drive
line to the qubit mode $i$.

We need to now compute the spectral densities of the baths corresponding
to the transmission lines. $\bar{\mathbf{C}}_{Z}\left(\omega\right)$
matrix defined in Eq. \eqref{eq:CZ-bar} is also a scalar in the case
of a single bath corresponding to the $d$-th drive line and is given
by

\begin{equation}
\bar{C}_{Z,d}\left(\omega\right)=-\frac{i\omega Z_{0}}{1+i\omega C_{p(d)}Z_{0}}
\end{equation}
Kernel of the bath due to the $d$-th drive line is given in Eq. (35)
of \cite{Burkard} as

\begin{equation}
K_{d}\left(\omega\right)=\frac{\bar{C}_{Z,d}\left(\omega\right)}{1+\bar{\mathbf{m}}_{d}^{T}\mathbf{C}^{-1}\bar{\mathbf{m}}_{d}\bar{C}_{Z,d}\left(\omega\right)}
\end{equation}
The term $\bar{\mathbf{m}}_{d}^{T}\mathbf{C}^{-1}\bar{\mathbf{m}}_{d}$
can be evaluated in the final frame using Eq. \eqref{eq:mij-final-frame}
and noting that $\mathbf{C=1}$ in the final frame. Hence

\begin{equation}
\bar{\mathbf{m}}_{d}^{T}\mathbf{C}^{-1}\bar{\mathbf{m}}_{d}=C_{p(d)}^{2}\underset{i}{\sum}\left(\mathrm{Im}\left[Z_{i,p(d)}\left(\omega_{i}\right)\right]\right)^{2}/L_{i}
\end{equation}

The spectrum of the bath is given by

\begin{eqnarray}
J_{d}\left(\omega\right) & = & -\mathrm{Im}\left[K_{d}\left(\omega\right)\right]\nonumber \\
 & = & \frac{\omega Z_{0}}{1+\omega^{2}Z_{0}^{2}\left(C_{p(d)}+\bar{\mathbf{m}}_{d}^{T}\mathbf{C}^{-1}\bar{\mathbf{m}}_{d}\right)^{2}}\nonumber \\
 & \simeq & \frac{\omega Z_{0}}{1+\omega^{2}Z_{0}^{2}C_{p(d)}^{2}}
\end{eqnarray}
assuming $\left(\bar{\mathbf{m}}_{j}^{T}\mathbf{C}^{-1}\bar{\mathbf{m}}_{j}\right)/C_{p(d)}\ll1$
which holds for typical parameter values and frequencies.

Finally $T_{1}$ rate of the qubit mode $i$ due to the $d$-th drive
line can be calculated using Eq. (44) of \cite{Burkard}

\begin{eqnarray}
\frac{1}{T_{1}^{i,d}} & = & \frac{4}{\hbar}\left|\left\langle 0\right|m_{id}\mathbf{\hat{Q}}_{i}\left|1_{i}\right\rangle \right|^{2}J_{d}\left(\omega_{i}\right)\coth(\frac{\hbar\omega_{i}}{2k_{B}T})
\end{eqnarray}
which can be simplified assuming $\coth(\frac{\hbar\omega_{i}}{2k_{B}T})\simeq1$
for the typical chip temperatures as

\begin{align}
\frac{1}{T_{1}^{i,d}} & =\frac{4}{\hbar}\left|\left\langle 0\right|m_{id}\mathbf{\hat{Q}}_{i}\left|1_{i}\right\rangle \right|^{2}J_{d}\left(\omega_{i}\right)\nonumber \\
 & =\frac{2}{L_{i}}\mathrm{Im}\left[Z_{i,p(d)}\left(\omega_{i}\right)\right]^{2}\frac{\omega_{i}^{2}Z_{0}C_{p(d)}^{2}}{1+\omega_{i}^{2}Z_{0}^{2}C_{p(d)}^{2}}\label{eq:T1-exp-long-1}
\end{align}

\begin{figure}
\includegraphics[scale=0.6]{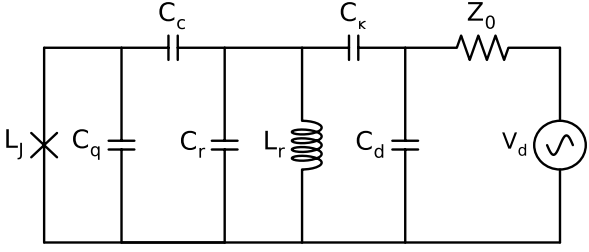}

\caption{\label{fig:Transmon-readout-line}Example circuit of a Transmon qubit
coupled to a readout resonator which in turn coupled to external drive
line of characteristic impedance $Z_{0}$. $C_{d}$ is the shunt capacitance
of the drive port.}
\end{figure}

To see that Purcell rates $\frac{1}{T_{1}^{i,d}}$'s are independent
of the drive port shunt capacitances $C_{p(d)}$'s we workout $Z_{i,p(d)}(\omega_{i})$
for the example circuit shown in Fig. \eqref{fig:Transmon-readout-line}.
Assuming $C_{r}\gg C_{\kappa}$, $C_{q}\gg C_{c}$ and $C_{d}\gg C_{\kappa}$
one can show that

\begin{equation}
\mathrm{Im}\left[Z_{12}(\omega)\right]\simeq\frac{\left(\frac{C_{c}C_{\kappa}}{C_{q}C_{d}}\right)L_{r}\omega}{1-(\omega/\omega_{r})^{2}}\label{eq:Z12-example}
\end{equation}
with $\omega_{r}=1/\sqrt{L_{r}C_{r}}$ and port $1$ being defined
across the Josephson junction and port $2$ across $C_{d}$. So that
the Purcell rate $\frac{1}{T_{1}^{i,j}}$ derived in Eq. \eqref{eq:T1-exp-long-1}
gives

\begin{align}
1/T_{1} & =\frac{2}{L_{q}}\mathrm{Im}\left[Z_{12}\left(\omega_{q}\right)\right]^{2}\frac{\omega_{q}^{2}Z_{0}C_{d}^{2}}{1+\omega_{q}^{2}Z_{0}^{2}C_{d}^{2}}\nonumber \\
 & =\frac{2}{L_{q}}\left(\frac{C_{c}C_{\kappa}}{C_{q}C_{r}}\right)^{2}\frac{(\omega_{q}/\omega_{r})^{4}}{\left[1-(\omega_{q}/\omega_{r})^{2}\right]^{2}}\left(\frac{Z_{0}}{1+\omega_{q}^{2}Z_{0}^{2}C_{d}^{2}}\right)\label{eq:T1-Cd}
\end{align}
where $\omega_{q}$ is the qubit frequency and $L_{q}$ qubit inductance.
The expression in the Eq. \eqref{eq:T1-Cd} above will be independent
of $C_{d}$, the total shunt capacitance of the drive port, in the
limit of $\omega_{d}^{2}Z_{0}^{2}C_{d}^{2}\ll1$ which holds for typical
parameter values in the actual experiments.

One can similarly calculate coupling of the qubit to the voltage source
$V_{d}$ using Eqs. \eqref{eq:eps-ij} and \eqref{eq:Z12-example}
to get

\begin{align}
\varepsilon_{12} & =e^{i\theta_{2}}\sqrt{\frac{\omega_{q}}{2\hbar L_{q}}}\mathrm{Im}\left[Z_{12}(\omega_{q})\right]C_{d}\nonumber \\
 & =\frac{1}{\sqrt{2\hbar Z_{q}}}\left(\frac{C_{c}C_{\kappa}}{C_{q}C_{r}}\right)\frac{(\omega_{q}/\omega_{r})^{2}}{1-(\omega_{q}/\omega_{r})^{2}}\left(\frac{e^{i\theta_{2}}}{\sqrt{1+\omega_{d}^{2}Z_{0}^{2}C_{d}^{2}}}\right)
\end{align}
where $\theta_{2}=\frac{\pi}{2}-\arctan(\omega_{d}Z_{0}C_{d})$ and
$Z_{q}=\sqrt{L_{q}/C_{q}}$. Above expression for the coupling $\varepsilon_{12}$
of the qubit to its voltage source $V_{d}$  will be again independent
of $C_{d}$ in the limit of $\omega_{d}^{2}Z_{0}^{2}C_{d}^{2}\ll1$
which holds for typical parameter values.

\section{\label{sec:Anharmonicities-Chi-shifts}Expressions for the Qubit
Anharmonicities and the Dispersive Shifts in the Resonator Frequencies}

In this section we derive expressions for the anharmonicity $\delta_{i}$
of the qubit mode $i$ and dispersive shift $\chi_{ik}$ in the frequency
$\omega_{R_{k}}$ of the resonator mode $k$ due to qubit mode $i$
using the results of Appendix \eqref{sub:Expansion-of-the-junction-potentials}.
Anharmonicities and dispersive shifts are generated by the nonlinear
terms in the expansion of the junction potentials.

From the term $H_{\beta}$ in Eq. \eqref{eq:H-beta} in the expansion
in Eq. \eqref{eq:Yale-normal-ordered-expansion} originally given
in \cite{BBQ-Yale} we note the following

\begin{eqnarray}
\delta_{i} & = & -12\beta_{iiii}\\
\chi_{ik} & = & -24\beta_{iikk}
\end{eqnarray}
Using the expression for $\beta_{pp'qq'}$ in Eq. \eqref{eq:beta-expression}
and Eqs. \eqref{eq:alpha-ii-exp} and \eqref{eq:alpha-ik-exp} we
obtain

\begin{eqnarray}
\delta_{i} & = & -E_{C}^{(i)}\left(\frac{\omega_{J_{i}}}{\omega_{i}}\right)^{2}\label{eq:lambda-anharmonicity}\\
\chi_{ik} & = & -2E_{C}^{(i)}\left(\frac{\omega_{J_{i}}^{2}}{\omega_{i}\omega_{R_{k}}}\right)r_{ki}^{2}C_{i}\left(\frac{\omega_{R_{k}}^{2}}{\omega_{R_{k}}^{2}-\omega_{i}^{2}}\right)^{2}
\end{eqnarray}
From Eq. \eqref{eq:Capacitance-matrix} we note

\begin{equation}
r_{ik}=\frac{2g_{ik}}{\sqrt{C_{i}\omega_{i}\omega_{R_{k}}}}
\end{equation}
Hence

\begin{eqnarray}
\chi_{ik} & = & -2E_{C}^{(i)}\left(\frac{\omega_{J_{i}}^{2}}{\omega_{i}\omega_{R_{k}}}\right)\left(\frac{4g_{ik}^{2}}{\omega_{i}\omega_{R_{k}}}\right)\left(\frac{\omega_{R_{k}}^{2}}{\omega_{R_{k}}^{2}-\omega_{i}^{2}}\right)^{2}\nonumber \\
 & = & 8\delta_{i}\left(\frac{g_{ik}\omega_{R_{k}}}{\omega_{R_{k}}^{2}-\omega_{i}^{2}}\right)^{2}\label{eq:chi-shift}
\end{eqnarray}

\section{Conclusion \& Outlook}

We have analyzed superconducting quantum processors consisting of
low anharmonicity transmon qubits. We have shown that the exchange
coupling rates between qubits is related in a simple way to the off-diagonal
entry of the multiport impedance matrix connecting the qubit ports
evaluated at qubit frequencies. Qubit ports are defined across the
Josephson junctions. Similarly coupling of the qubits to their drives
and Purcell relaxation rates of the qubit modes are related to the
entry of the multiport impedance matrix connecting the qubits and
the drive ports. This gives a complete microwave description of the
system in the qubit subspace. The formulas requiring only evaluation
at qubit frequencies(no need for frequency sweeps and fitting) make
modeling and simulation of the chips much more efficient.

Simple relations of the qubit exchange coupling rates and the couplings
of the qubits to the voltage drives to the impedance response allow
application of microwave engineering techniques to improve the performance
of the two-qubit gates. One application could be to use microwave
coupler or filtering structures to shape the response profile to reduce
unwanted terms in two-qubit gates.

\section{Acknowledgements}

We thank Easwar Magesan and Hanhee Paik for useful discussions and
Salvatore Olivadese for support with microwave simulations. DD acknowledges
support from Intelligence Advanced Research Projects Activity (IARPA)
under contract W911NF-16-0114.

\clearpage{}

\section{Appendix}

\subsection{\label{sub:Derivation-of-the-Hamiltonian}Derivation of the Hamiltonian
for the Canonical Multiport Cauer Circuit}

Any multiport lossless impedance response can be synthesized with
the canonical Cauer circuit shown in Fig. \eqref{fig:Cauer-circuit}.
The Cauer circuit consists of $N$ ``qubit ports'' on the left shunted
by the Josephson junctions in our case and $M$ internal modes synthesized
as parallel $LC$ tank circuits on the right. Couplings between the
ports and the internal modes are mediated by the multiport Belevitch
transformers (see \cite{Newcomb} for details). A purely capacitive
stage (upper right) provides total shunt capacitances of the junctions.
In the most general form shown in Fig. \eqref{fig:Cauer-circuit}
there is a purely inductive stage shown in the lower right corner.
This stage is responsible of the purely inductive energy storage in
the system. However in most of the physical situations arising with
distributed electromagnetic structures this stage will be absent since
any distributed inductor will always have a finite parasitic capacitance.
For cases where such a stage is really necessary the degrees of freedoms
associated with it can be eliminated with a Born-Oppenheimer analysis
\cite{Brito}.

The synthesis of the canonical Cauer circuit in Fig. \eqref{fig:Cauer-circuit}
proceeds as follows: first we do the eigendecomposition of the residue
at DC $\mathbf{A}_{0}$ in Eq. \eqref{eq:Partial-Fraction-Expansion}
\begin{equation}
\mathbf{A}_{0}=\mathbf{U}\mathbf{C}_{0}^{-1}\mathbf{U}^{T}\label{eq:A0-C0}
\end{equation}
where $\mathbf{U}$ is the $N\times N$ orthonormal matrix holding
the eigenvectors of $\mathbf{A}_{0}$ and $\mathbf{C}_{0}$ is the
diagonal matrix with entries $\left(C_{1},\ldots,C_{N}\right)$, inverses
of eigenvalues of $\mathbf{A}_{0}$. Entries of $\mathbf{U}$ are
the turns ratios of the multiport Belevitch transformer corresponding
to the purely capacitive stage in Fig. \eqref{fig:Cauer-circuit}.
In the case of no direct electrostatic interaction between the qubit
port terminals $\mathbf{U}$ will be simply the identity matrix.

For the internal modes of frequency $\omega_{R_{k}}=1/\sqrt{L_{R_{k}}C_{R_{k}}}$
we choose a characteristic impedance of $Z_{0}=1/\omega_{R_{k}}$
that will make all $C_{R_{k}}=1$ for $1\leq k\leq M$. There is a
freedom in the choice of this characteristic impedance; this choice
should have no effect on the physical coupling rates. With this choice
we have $L_{R_{k}}=1/\omega_{R_{k}}^{2}$. Then with $\mathbf{A}_{k}$'s
being rank-1 matrices \cite{DDV} and with our choice of $C_{R_{k}}=1$

\begin{equation}
\mathbf{A}_{k}=r_{k}^{T}r_{k}
\end{equation}
where $r_{k}$ is the row-vector $r_{k}=(r_{k1},\ldots,r_{kN})$ for
$1\leq k\leq M$. $r_{k}$'s constitute rows of turn ratios of the
multiport Belevitch transformer matrix $\mathbf{R}$ connecting the
internal modes to the ports.

The final purely inductive stage can be synthesized in a similar way
to the purely capacitive DC stage with a eigendecomposition of the
$\mathbf{A}_{\infty}$ matrix

\begin{equation}
\mathbf{A}_{\infty}=\mathbf{T}^{T}\mathbf{L}_{\infty}\mathbf{T}
\end{equation}
with $\mathbf{T}$ being the orthonormal matrix holding the eigenvectors
and the diagonal matrix $\mathbf{L}_{\infty}$ holding the inductances
$(L_{1}^{\infty},\ldots,L_{N}^{\infty})$.

Using the lumped element circuit quantization method in \cite{Burkard}
together with a technique to handle multiport Belevitch transformers
\cite{Solgun} we can identify the degrees of freedom in the Cauer
circuit in Fig. \eqref{fig:Cauer-circuit} and write an equation of
motion. The effective fundamental loop matrix defined in Eq. (21)
of \cite{Burkard} is

\begin{equation}
\mathbf{F}_{C}=\left(\begin{array}{c}
\mathbf{U}\\
-\mathbf{R}\mathbf{U}
\end{array}\right)
\end{equation}

The Hamiltonian is

\begin{equation}
\mathcal{H}=\frac{1}{2}\mathbf{Q}^{T}\mathbf{C}^{-1}\mathbf{Q}+\frac{1}{2}\mathbf{\Phi}^{T}\mathbf{M}_{0}\mathbf{\Phi}-\stackrel[i=1]{N}{\sum}E_{J_{i}}\cos\left(\varphi_{J_{i}}\right)\label{eq:Cauer-Hamiltonian}
\end{equation}
where $\mathbf{\Phi}=(\Phi_{J_{1}},\ldots,\Phi_{J_{N}},\Phi_{R_{1}},\ldots,\Phi_{R_{M}})^{T}$
being the flux coordinate vector. $\varphi_{J_{i}}$ is the phase
of the junction $i$ related to the flux across it by the Josephson
relation $\Phi_{J_{i}}=\frac{\mathrm{\Phi}_{0}}{2\pi}\varphi_{J_{i}}$,
for $1\leq i\leq N$. $\Phi_{R_{k}}$ is the flux across the inductor
of the internal mode $k,$ $1\leq k\leq M$. $E_{J_{i}}$ is the Josephson
energy of junction $i$ related to its inductance $L_{J_{i}}$ by
$E_{J_{i}}=\left(\frac{\mathrm{\Phi}_{0}}{2\pi}\right)^{2}\frac{1}{L_{J_{i}}}$.
The capacitance matrix $\mathbf{C}$ is given by

\begin{align}
\mathbf{C} & =\mathbf{F}_{C}\mathbf{C}_{0}\mathbf{F}_{C}^{T}+\mathbf{C}_{R}\nonumber \\
 & =\left(\begin{array}{cc}
\mathbf{U}\mathbf{C}_{0}\mathbf{U}^{T} & -\mathbf{U}\mathbf{C}_{0}\mathbf{U}^{T}\mathbf{R}^{T}\\
-\mathbf{R}\mathbf{U}\mathbf{C}_{0}\mathbf{U}^{T} & \mathbf{C}_{R}+\mathbf{R}\mathbf{U}\mathbf{C}_{0}\mathbf{U}^{T}\mathbf{R}^{T}
\end{array}\right)
\end{align}
The capacitance matrix becomes

\begin{equation}
\mathbf{C}=\left(\begin{array}{cc}
\mathbf{C}_{0} & -\mathbf{C}_{0}\mathbf{R}^{T}\\
-\mathbf{R}\mathbf{C}_{0} & \mathbf{1}_{M\times M}+\mathbf{R}\mathbf{C}_{0}\mathbf{R}^{T}
\end{array}\right)
\end{equation}
in the absence of direct electrostatic dipole interactions between
the ports since $\mathbf{U}$ is the identity matrix in that case
and with our choice of $\mathbf{C}_{R}=\mathbf{1}_{M\times M}$ for
the capacitances of the internal modes.

$\mathbf{M}_{0}$ is the diagonal matrix holding the inverses of the
inductances of the internal modes on its diagonal

\begin{equation}
\mathbf{M}_{0}=\left(\begin{array}{cccc}
\mathbf{0}_{N\times N} &  &  & \mathbf{0}\\
 & 1/L_{R_{1}}\\
 &  & \ddots\\
\mathbf{0} &  &  & 1/L_{R_{M}}
\end{array}\right)
\end{equation}

\afterpage{%
\onecolumngrid

\begin{figure}
\begin{centering}
\includegraphics[scale=0.8]{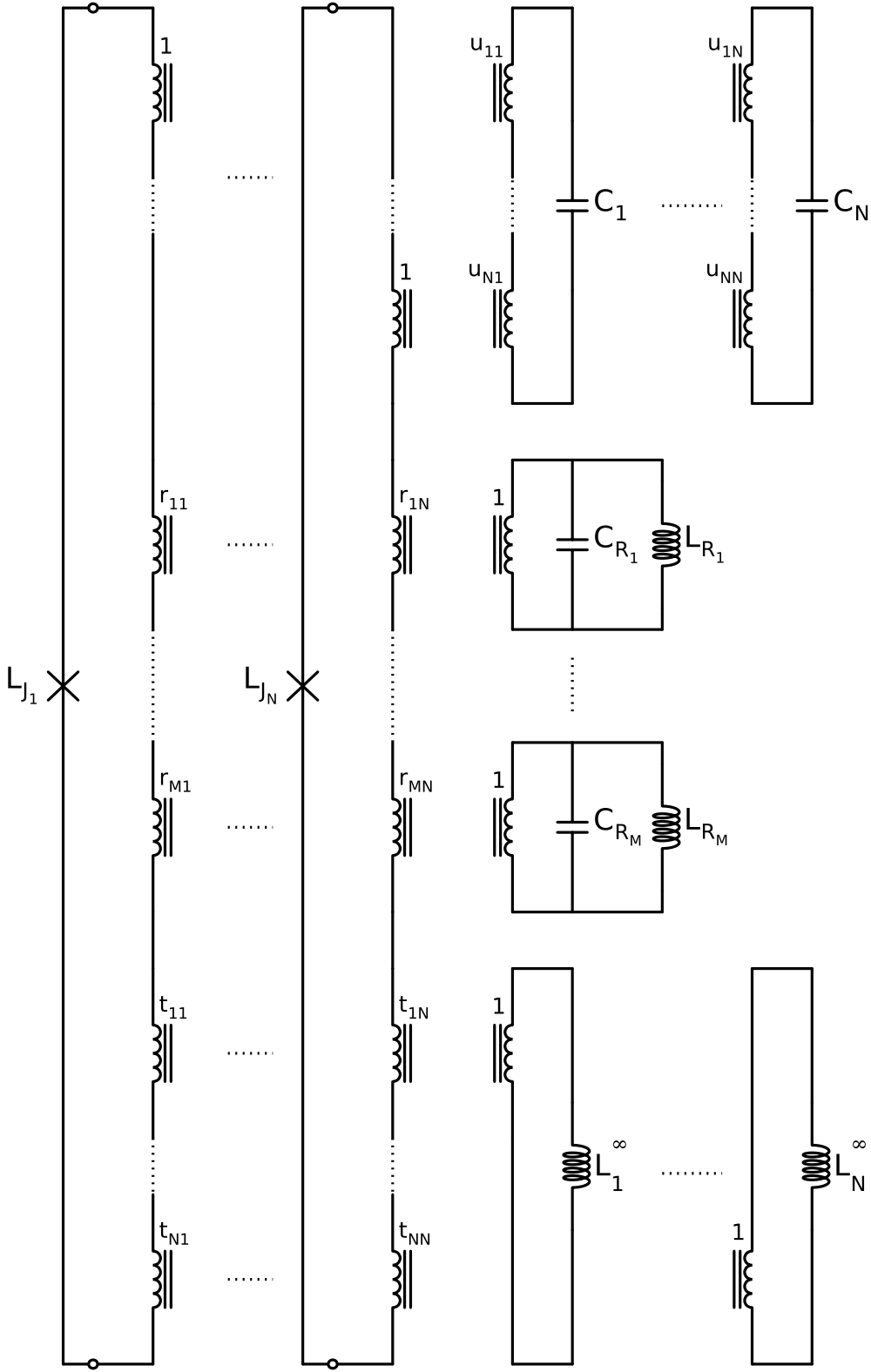}
\par\end{centering}

\caption{\label{fig:Cauer-circuit}Multiport canonical Cauer circuit shunted
with Josephson junctions. On the left we have $N$ ``qubit ports''
shunted by the Josephson junctions $J_{i}$'s. Port terminals are
shown with open dots. On the upper-right we have the purely capacitive
stage providing the total qubit shunt capacitances $C_{i}'s$. This
stage is coupled to the rest of the circuit with the multiport Belevitch
transformer $\mathbf{U}$. $LC$ tank circuits on the middle right
correspond to the internal modes. Interactions between the qubit ports
and the internal modes are mediated by the multiport Belevitch transformer
of turns ratio matrix $\mathbf{R}$. On the lower right we have the
purely inductive stage corresponding to the pole at infinity of the
impedance expansion in Eq. \eqref{eq:Partial-Fraction-Expansion}
consisting of linear inductors $L_{1}^{\infty},\ldots,L_{N}^{\infty}$.
This stage can safely be neglected since in most of the physical situations
every inductor will have a parasitic capacitance.}
\end{figure}
\clearpage
\twocolumngrid
}

\clearpage{}

\subsection{\label{sub:Derivation-of-the-classical-drive-cross-talk-appendix}\textcolor{black}{Derivation
of the Couplings Rates of the Qubits to the Voltage Drives}}

In this appendix we augment the canonical Cauer circuit in Fig. \eqref{fig:Cauer-circuit}
by including the drive lines as shown in Fig. \eqref{fig:Cauer-circuit-with-drives}.
We added $N_{D}$ drive lines hence $N_{D}$ more ports. Drive line
for the qubit $i$ consists of the voltage source $V_{d(i)}$ driving
the transmission line of characteristic impedance $Z_{0}$ whose other
end is connected to the drive port $d(i)$ (Here we are assuming that
$d(i)$ is the index number of the drive port corresponding to the
qubit $i$). Synthesis of such a circuit from an impedance matrix
$\mathbf{Z}(\omega)$ proceeds as described in the previous section,
this time with $N+N_{D}$ ports.

Again using the method in \cite{Burkard} we obtain the following
Hamiltonian for the augmented Cauer circuit in Fig. \eqref{fig:Cauer-circuit-with-drives}

\begin{widetext}

\begin{equation}
\mathcal{H}=\frac{1}{2}(\mathbf{Q}-\mathbf{C}_{Q}\ast\mathbf{V}_{V})^{T}\mathbf{C}^{-1}(\mathbf{Q}-\mathbf{C}_{Q}\ast\mathbf{V}_{V})+\frac{1}{2}\mathbf{\Phi}^{T}\mathbf{M}_{0}\mathbf{\Phi}-\stackrel[i=1]{N}{\sum}E_{J_{i}}\cos\left(\varphi_{J_{i}}\right)\label{eq:Augmented-Cauer-Hamiltonian}
\end{equation}
\end{widetext}\textcolor{black}{where as in the previous section}
$\mathbf{\Phi}=\left(\Phi_{J_{1}},\ldots,\Phi_{J_{N}},\Phi_{R_{1}},\ldots,\Phi_{R_{M}}\right)^{T}$
is the flux coordinate vector. The capacitance matrix $\mathbf{C}$
is given by
\begin{equation}
\mathbf{C}=\mathbf{F}_{C}\mathbf{C}_{S}\mathbf{F}_{C}^{T}+\mathbf{C}_{R}
\end{equation}
where $\mathbf{C}_{S}$ is the diagonal matrix holding the total shunt
capacitances seen at the ports

\begin{equation}
\mathbf{C}_{S}=\left(\begin{array}{cc}
\mathbf{C}_{0} & \mathbf{0}\\
\mathbf{0} & \mathbf{C}_{D}
\end{array}\right)\label{eq:Cs-matrix}
\end{equation}
where $\mathbf{C}_{0}$ and $\mathbf{C}_{D}$ are $N\times N$ and
$N_{D}\times N_{D}$ diagonal matrices holding total capacitances
shunting the qubit and drive ports, respectively such that

\begin{equation}
\mathbf{C}_{0}=\left(\begin{array}{ccc}
C_{1} &  & \mathbf{0}\\
 & \ddots\\
\mathbf{0} &  & C_{N}
\end{array}\right)
\end{equation}

\begin{equation}
\mathbf{C}_{D}=\left(\begin{array}{ccc}
C_{N+1} &  & \mathbf{0}\\
 & \ddots\\
\mathbf{0} &  & C_{N+N_{D}}
\end{array}\right)\label{eq:Cd-def}
\end{equation}
where capacitances $(C_{1},\ldots,C_{N},C_{N+1},\ldots,C_{N+N_{D}})$
are shown in Fig.\ref{fig:Cauer-circuit-with-drives} in the purely
capacitive stage coupled to the rest of the circuit with the multiport
Belevitch transformer $\mathbf{U}$. $\mathbf{C}_{R}$ is again the
identity matrix in the resonator subspace and neglecting any electrostatic
dipole interaction between the ports (i.e. $\mathbf{U}=\mathbf{1}$)
the fundamental loop matrix $\mathbf{F}_{C}$ is given by

\begin{equation}
\mathbf{F}_{C}=\left(\begin{array}{cc}
\mathbf{1}_{N\times N} & \mathbf{0}_{N\times N_{D}}\\
-\mathbf{R} & -\mathbf{V}
\end{array}\right)\label{eq:FC-drives}
\end{equation}
where $\mathbf{F}_{C}$, $\mathbf{R}$ and $\mathbf{V}$ are $(N+M)\times(N+N_{D})$
, $(M\times N)$ and $(M\times N_{D})$ matrices, respectively. Matrices
$\mathbf{R}$ and $\mathbf{V}$ are multiport Belevitch transformer
matrices (with turn ratio entries $r_{ki}$ and $v_{kd}$ as shown
in Fig. \eqref{fig:Cauer-circuit-with-drives} for $1\leq k\leq M$,
$1\leq i\leq N$ and $1\leq d\leq N_{D}$) mediating the couplings
of the internal modes to the qubits and the voltage sources, respectively.
The diagonal matrix $\mathbf{M}_{0}$ again holds the inverses of
the inductances of the internal modes on its diagonal

\begin{equation}
\mathbf{M}_{0}=\left(\begin{array}{cccc}
\mathbf{0}_{N\times N} &  &  & \mathbf{0}\\
 & 1/L_{R_{1}}\\
 &  & \ddots\\
\mathbf{0} &  &  & 1/L_{R_{M}}
\end{array}\right)
\end{equation}
\textcolor{black}{$\mathbf{V}_{V}=(V_{1},\ldots,V_{N_{D}})$ is the
vector of voltage sources and $\ast$ is the time convolution operator.
$\mathbf{C}_{Q}$ is the $(N+M)\times N_{D}$ matrix coupling the
voltage source $\mathbf{V}_{V}$ vector to the charge coordinates
$\mathbf{Q}$ and is given by}

\textcolor{black}{
\begin{equation}
\mathbf{C}_{Q}=\mathrm{\mathbf{C}}_{V}+\mathcal{C}_{V}
\end{equation}
As we will show below $\mathrm{\mathbf{C}}_{V}$ is frequency independent
whereas $\mathcal{C}_{V}$ is non-zero only for AC voltage drives.
$\mathrm{\mathbf{C}}_{V}$ is given in Eq. (23) in \cite{Burkard}
as}

\begin{equation}
\mathrm{\mathbf{C}}_{V}=\mathbf{F}_{C}\mathbf{C}_{S}\mathbf{F}_{VC}^{T}\label{eq:CV-DC-part}
\end{equation}
where the loop matrix $\mathbf{F}_{VC}$ is given by

\begin{equation}
\mathbf{F}_{VC}=\left(\begin{array}{cc}
\mathbf{0}_{N\times N} & \mathbf{1}_{N\times N_{D}}\end{array}\right)\label{eq:FVC-FZC-kappa-no-direct-dipole-assumption}
\end{equation}

$\mathcal{C}_{V}(\omega)$ is given in Eq. (7.25) of \cite{Solgun}
which is an extension of \textcolor{black}{Eq. (23) of \cite{Burkard}}
to AC voltage sources

\begin{equation}
\mathcal{C}_{V}(\omega)=\bar{\mathbf{m}}\bar{\mathbf{C}}_{Z}\bar{\mathbf{m}}_{V}^{T}\label{eq:CV}
\end{equation}
where from Eqs. (7.19-7.21) in \cite{Solgun}

\begin{eqnarray}
\bar{\mathbf{m}} & = & \mathbf{F}_{C}\mathbf{C}_{S}\mathbf{F}_{ZC}^{T}\label{eq:mbar}\\
\bar{\mathbf{m}}_{V} & = & \mathbf{F}_{VC}\mathbf{C}_{S}\mathbf{F}_{ZC}^{T}\\
\bar{\mathbf{C}}_{Z}(\omega) & = & -i\omega\mathbf{Z}_{0}\left[\mathbf{1}+i\omega\mathbf{F}_{ZC}\mathbf{C}_{S}\mathbf{F}_{ZC}^{T}\mathbf{Z}_{0}\right]^{-1}\label{eq:CZ-bar}
\end{eqnarray}
Here $\mathbf{F}_{ZC}=\mathbf{F}_{VC}$ given in Eq. \eqref{eq:FVC-FZC-kappa-no-direct-dipole-assumption}
and $\mathbf{Z}_{0}$ is the $N_{D}\times N_{D}$ matrix giving the
multiport impedance seen at the drive ports looking into the environment
away from the chip and is simply the diagonal matrix consisting of
diagonal entries $Z_{0}$'s.

\textcolor{black}{We observe that $\mathrm{\mathbf{C}_{V}=\bar{\mathbf{m}}}$
since $\mathbf{F}_{VC}=\mathbf{F}_{ZC}$. Noting}

\begin{equation}
\mathbf{F}_{ZC}\mathbf{C}_{S}\mathbf{F}_{ZC}^{T}=\mathbf{C}_{D}\label{eq:Ck-def}
\end{equation}
we write

\begin{equation}
\bar{\mathbf{C}}_{Z}(\omega)=-i\omega\mathbf{Z}_{0}\left[\mathbf{1}+i\omega\mathbf{C}_{D}\mathbf{Z}_{0}\right]^{-1}
\end{equation}

We now work out $\bar{\mathbf{m}}$ using Eqs. \eqref{eq:mbar}, \eqref{eq:Cs-matrix},
\eqref{eq:FC-drives}, \eqref{eq:FVC-FZC-kappa-no-direct-dipole-assumption}
and noting $\mathbf{F}_{ZC}=\mathbf{F}_{VC}$ 

\begin{eqnarray}
\bar{\mathbf{m}} & = & \mathbf{F}_{C}\mathbf{C}_{S}\mathbf{F}_{ZC}^{T}\nonumber \\
 & = & \left(\begin{array}{cc}
\mathbf{1} & \mathbf{0}\\
-\mathbf{R} & -\mathbf{V}
\end{array}\right)\left(\begin{array}{cc}
\mathbf{C}_{0} & \mathbf{0}\\
\mathbf{0} & \mathbf{C}_{D}
\end{array}\right)\left(\begin{array}{cc}
\mathbf{0} & \mathbf{1}\end{array}\right)^{T}\nonumber \\
 & = & \left(\begin{array}{c}
\mathbf{0}\\
-\mathbf{V}\mathbf{C}_{D}
\end{array}\right)\label{eq:mbar-worked-out}
\end{eqnarray}
Applying the capacitance rescaling $\mathbf{\Phi}_{J}\rightarrow\mathbf{C}_{0}^{1/2}\mathbf{\Phi}_{J}$
and the transformation in Eq. \eqref{eq:T-transformation} to $\bar{\mathbf{m}}$
in Eq. \eqref{eq:mbar-worked-out} 

\begin{equation}
\bar{\mathbf{m}}\rightarrow\mathbf{T}^{t}\left(\begin{array}{cc}
\mathbf{C}_{0}^{-1/2} & \mathbf{0}\\
\mathbf{0} & \mathbf{1}
\end{array}\right)\bar{\mathbf{m}}
\end{equation}
we get

\begin{eqnarray}
\bar{\mathbf{m}} & = & \left(\begin{array}{cc}
\mathbf{1} & \mathbf{0}\\
\mathbf{R}\mathbf{C}_{0}^{1/2} & \mathbf{1}
\end{array}\right)\left(\begin{array}{cc}
\mathbf{C}_{0}^{-1/2} & \mathbf{0}\\
\mathbf{0} & \mathbf{1}
\end{array}\right)\left(\begin{array}{c}
\mathbf{0}\\
-\mathbf{V}
\end{array}\right)\mathbf{C}_{D}\nonumber \\
 & = & \left(\begin{array}{c}
\mathbf{0}\\
-\mathbf{V}\mathbf{C}_{D}
\end{array}\right)\label{eq:mbar-before-Schr-Wollf}
\end{eqnarray}
We note that $\bar{\mathbf{m}}$ is unaffected by this transformation.
\textcolor{black}{Since $\mathrm{\mathbf{C}_{V}=\bar{\mathbf{m}}}$
we have after the transformations}

\begin{equation}
\mathrm{\mathbf{C}}_{V}=\left(\begin{array}{c}
\mathbf{0}\\
-\mathbf{V\mathbf{C}}_{D}
\end{array}\right)
\end{equation}

Noting

\begin{eqnarray}
\bar{\mathbf{m}}_{V} & = & \mathbf{F}_{VC}\mathbf{C}\mathbf{F}_{ZC}^{T}\nonumber \\
 & = & \mathbf{F}_{ZC}\mathbf{C}\mathbf{F}_{ZC}^{T}\nonumber \\
 & = & \mathbf{C}_{D}
\end{eqnarray}
 we can write Eq. \eqref{eq:CV} as

\begin{eqnarray}
\mathcal{C}_{V}(\omega) & = & \bar{\mathbf{m}}\bar{\mathbf{C}}_{Z}\bar{\mathbf{m}}_{V}^{T}\nonumber \\
 & = & \left(\begin{array}{c}
\mathbf{0}\\
\mathbf{V}
\end{array}\right)\widetilde{\mathbf{C}}_{Z}(\omega)
\end{eqnarray}
where we defined the $N_{D}\times N_{D}$ matrix

\begin{equation}
\widetilde{\mathbf{C}}_{Z}\left(\omega\right)=i\omega\mathbf{C}_{D}\mathbf{Z}_{0}\left[\mathbf{1}+i\omega\mathbf{C}_{D}\mathbf{Z}_{0}\right]^{-1}\mathbf{C}_{D}
\end{equation}

We have one final step to do, that is to apply the Schrieffer-Wolff
transformation \textcolor{black}{to $\mathrm{\mathbf{C}}_{V}$ and
}$\mathcal{C}_{V}(\omega)$ such that

\begin{eqnarray}
\mathrm{\mathbf{C}}_{V} & \rightarrow & \exp(-\mathbf{S})\mathrm{\mathbf{C}}_{V}\\
\mathcal{C}_{V}(\omega) & \rightarrow & \exp(-\mathbf{S})\mathcal{C}_{V}(\omega)
\end{eqnarray}
Using Eqs. (B.4) and (B.12a) of \cite{Winkler} and noting block structures
of matrices $\mathbf{S}$, $\mathrm{C}_{V}$ and $\mathcal{C}_{V}(\omega)$
we first define the following $(N+M)\times N_{D}$ matrix $\mathbf{D}$
having the $\left(i,d\right)$-th entry $D_{id}$ in the qubit subspace:

\begin{eqnarray}
D_{id} & = & \left[\exp(-\mathbf{S})\left(\begin{array}{c}
\mathbf{0}\\
-\mathbf{V}
\end{array}\right)\right]_{id}\nonumber \\
 & = & -\underset{k}{\sum}\left(\mathbf{M}_{1}\right)_{ik}\frac{v_{kd}}{\omega_{i}^{2}-\omega_{R_{k}}^{2}}\nonumber \\
 & =- & \omega_{i}^{2}C_{i}^{1/2}\underset{k}{\sum}\frac{r_{ki}v_{kd}}{\omega_{i}^{2}-\omega_{R_{k}}^{2}}\nonumber \\
 & = & -\omega_{i}^{2}C_{i}^{1/2}\underset{k}{\sum}\frac{[\mathbf{A}_{k}]_{i,p(d)}}{\omega_{i}^{2}-\omega_{R_{k}}^{2}}\nonumber \\
 & = & \omega_{i}C_{i}^{1/2}\mathrm{Im}\left[Z_{i,p(d)}(\omega_{i})\right]\nonumber \\
 & = & \mathrm{Im}\left[Z_{i,p(d)}\left(\omega_{i}\right)\right]/\sqrt{L_{i}}\label{eq:D_ij-equation}
\end{eqnarray}
where $v_{kd}$ is the $\left(k,d\right)$-th entry of $\mathbf{V}$
for $1\leq k\leq M$ and $1\leq d\leq N_{D}$. In the third line above
we used Eq. \eqref{eq:Mik} to replace $\left(\mathbf{M}_{1}\right)_{ik}$
with $\omega_{i}^{2}C_{i}^{1/2}r_{ki}$ and in the fourth line $[\mathbf{A}_{k}]_{i,p(d)}=r_{ki}v_{kd}$
where $[\mathbf{A}_{k}]_{i,p(d)}$ is the entry of the residue matrix
$\mathbf{A}_{k}$ in the impedance expansion in Eq. \eqref{eq:Partial-Fraction-Expansion}
for the circuit in Fig. \eqref{fig:Cauer-circuit-with-drives} connecting
the qubit port $i$ to drive port (with port index $p(d)$) corresponding
to the voltage source $V_{d}$. Hence $\mathbf{C}_{Q}$ transforms
to

\begin{equation}
\mathbf{C}_{Q}\rightarrow\mathbf{C}_{q}=\mathbf{D}\left(\mathbf{C}_{D}-\widetilde{\mathbf{C}}_{Z}(\omega)\right)\label{eq:Cq}
\end{equation}
Then one can write the following Hamiltonian in the final frame corresponding
to $\widetilde{\mathbf{M}}_{1}$ in Eq. \eqref{eq:block-diagonal-M1}

\begin{equation}
H=\frac{1}{2}\left(\boldsymbol{q}-\mathbf{C}_{q}*\mathbf{V}_{V}\right)^{T}\left(\boldsymbol{q}-\mathbf{C}_{q}*\mathbf{V}_{V}\right)+\frac{1}{2}\boldsymbol{\phi}^{T}\widetilde{\mathbf{M}}_{1}\boldsymbol{\phi}+\mathcal{O}(\boldsymbol{\varphi}_{J}^{4})\label{eq:Hamiltonian-drive-final-frame}
\end{equation}
with the $(N+M)\times N_{D}$ matrix $\mathbf{C}_{q}$ giving the
couplings of the voltage drives $\mathbf{V}_{V}$ to the momentum
degrees of freedom $\boldsymbol{q}$ in the final frame. After quantization
by introducing $\hat{q}_{i}=-i\sqrt{\frac{\hbar}{2Z_{i}}}(\hat{b}_{i}-\hat{b}_{i}^{\dagger})$
and noting that the characteristic impedance $Z_{i}$ of the qubit
mode $i$ is $Z_{i}=1/\omega_{i}$ we get the drive term on qubit
$i$ due to voltage source $V_{d}$

\begin{equation}
H_{id}^{D}=i\sqrt{\frac{\hbar\omega_{i}}{2}}\left[\mathbf{C}_{q}(\omega_{d})\right]_{i,d}V_{d}(\hat{b}_{i}-\hat{b}_{i}^{\dagger})
\end{equation}
where $\left[\mathbf{C}_{q}(\omega_{d})\right]_{i,d}$ is $(i,d)$-th
entry of $\mathbf{C}_{q}$ evaluated at the frequency $\omega_{d}$
(We assumed that $V_{d}$ is a single-tone sinusoidal voltage drive
at frequency $\omega_{d}$). In the case of zero off-chip crosstalk
$\widetilde{\mathbf{C}}_{Z}$ is diagonal and using Eqs. \eqref{eq:D_ij-equation}
and \eqref{eq:Cq} we have

\begin{eqnarray}
H_{id}^{D} & = & i\sqrt{\frac{\hbar\omega_{i}}{2L_{i}}}\mathrm{Im}\left[Z_{i,p(d)}(\omega_{i})\right]\frac{C_{p(d)}V_{d}(\hat{b}_{i}-\hat{b}_{i}^{\dagger})}{1+i\omega_{d}Z_{0}C_{p(d)}}
\end{eqnarray}
where $C_{p(d)}$ is the $d$-th diagonal entry of $\mathbf{C}_{D}$.
We note here that $\varepsilon_{id}$ in Eq. \eqref{eq:Hamiltonian}
is

\begin{equation}
\varepsilon_{id}=i\sqrt{\frac{\omega_{i}}{2\hbar L_{i}}}\mathrm{Im}\left[Z_{i,p(d)}(\omega_{i})\right]\left(\frac{C_{p(d)}}{1+i\omega_{d}Z_{0}C_{p(d)}}\right)
\end{equation}
which can be also written as

\begin{equation}
\varepsilon_{id}=\sqrt{\frac{\omega_{i}}{2\hbar L_{i}}}\mathrm{Im}\left[Z_{i,p(d)}(\omega_{i})\right]\frac{e^{i\theta_{d}}C_{p(d)}}{\sqrt{1+\omega_{d}^{2}Z_{0}^{2}C_{p(d)}^{2}}}
\end{equation}
with $\theta_{d}=\frac{\pi}{2}-\arctan(\omega_{d}Z_{0}C_{p(d)})$.

One can then define the following quantity (in units of $dB$) as
a measure of classical on-chip cross-talk on qubit $i$ while driving
qubit $j$

\begin{eqnarray}
X_{ij} & = & 20\mathrm{log}_{10}\left(\sqrt{\frac{\omega_{i}L_{J_{j}}}{\omega_{j}L_{J_{i}}}}\right)+20\mathrm{log}_{10}\left(\frac{\mathrm{Im}[Z_{i,d(j)}(\omega_{i})]}{\mathrm{Im}[Z_{j,d(j)}(\omega_{j})]}\right)\nonumber \\
 & \simeq & 20\mathrm{log}_{10}\left(\frac{\mathrm{Im}[Z_{i,d(j)}(\omega_{i})]}{\mathrm{Im}[Z_{j,d(j)}(\omega_{j})]}\right)
\end{eqnarray}
In the definition of the above crosstalk measure we neglected \textcolor{black}{the
term involving qubit frequencies} and junction inductances assuming
similar values.

\afterpage{%

\onecolumngrid

\begin{figure}
\begin{centering}
\includegraphics[angle=90,scale=0.75]{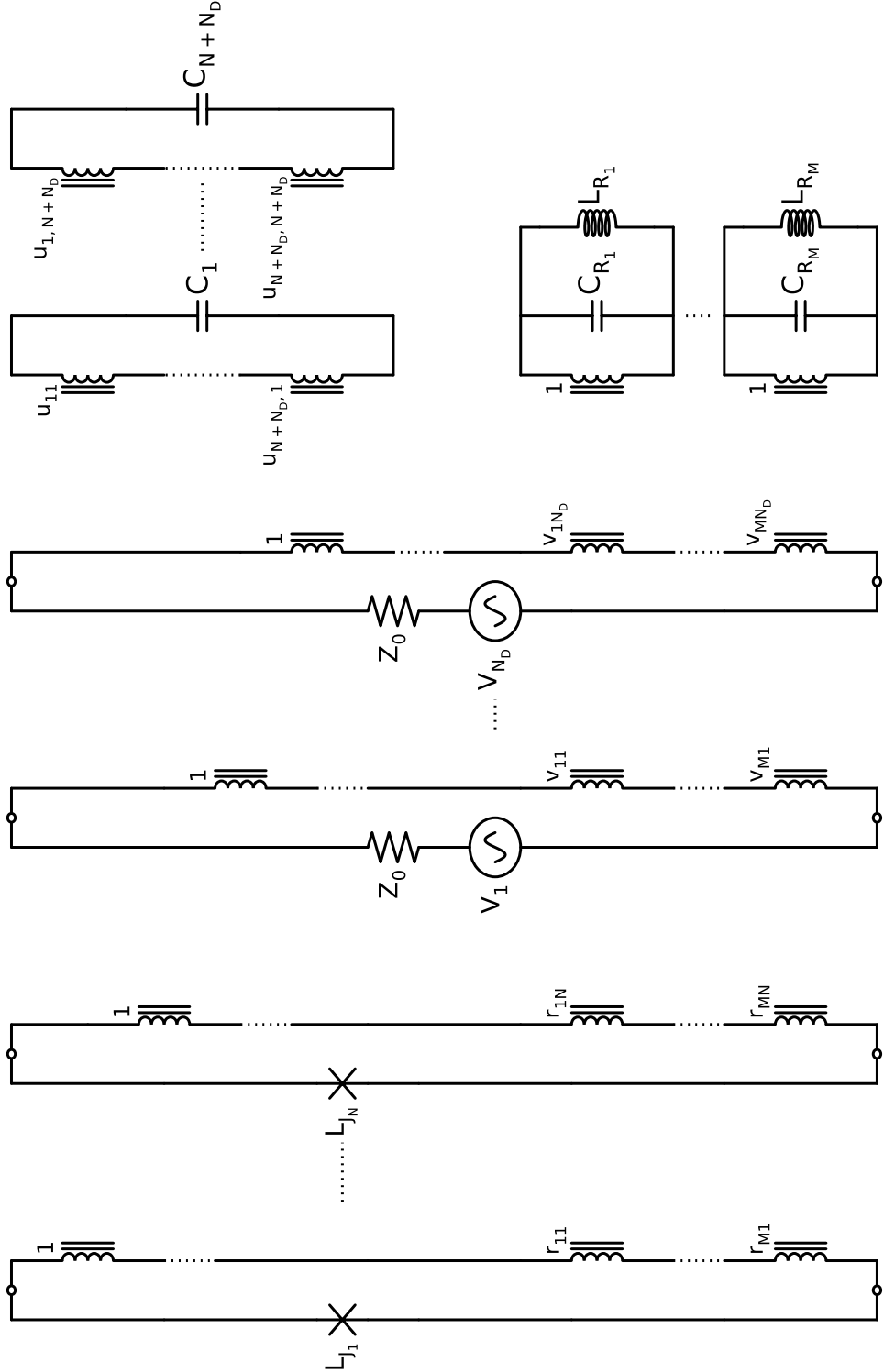}
\par\end{centering}

\caption{\label{fig:Cauer-circuit-with-drives}Canonical Cauer circuit in Fig.\ref{fig:Cauer-circuit}
with the purely inductive stage dropped and augmented with $N_{D}$
control/readout lines. Drive lines are modeled with transmission lines
of infinite extent such that they are represented by constant characteristic
impedances $Z_{0}$'s shunted with the voltage sources $V_{d}$'s
for $1\leq d\leq N_{D}$. Coupling of the drive lines to the internal
modes is mediated by the multiport Belevitch tranformer $\mathbf{V}$
with entries $v_{kd}$'s for $1\leq k\leq M$ and $1\leq d\leq N_{D}$.
The purely capacitive stage has now $N_{D}$ additional capacitances
$C_{N+1},\ldots,C_{N+N_{D}}$ corresponding to the shunt capacitances
of the drive ports (We define the $N_{D}\times N_{D}$ diagonal matrix
$\mathbf{C}_{D}$ in Eq.\ref{eq:Cd-def} holding total shunt capacitances
of the drive ports). The $(N+N_{D})\times(N+N_{D})$ multiport transformer
$\mathbf{U}$ with entries couples the shunt capacitances to the qubit
and drive ports.}
\end{figure}

\clearpage
\twocolumngrid
}

\clearpage{}

\afterpage{%
\onecolumngrid

\begin{figure}[H]
\includegraphics[scale=0.5]{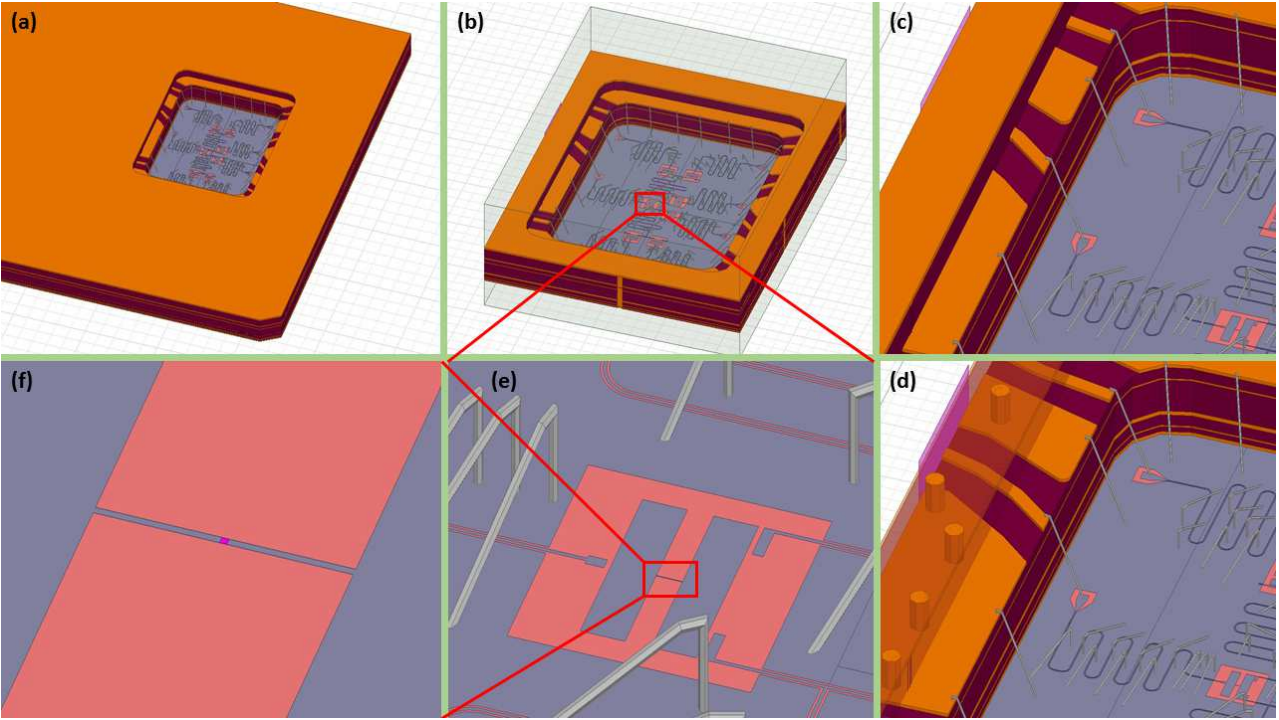}

\caption{\label{fig:Port-Definitions}a) HFSS \cite{HFSS} Model of a 7-Qubit
Device to illustrate the definition of the drive ports and the qubit
ports. The device consists of a superconducting quantum chip packaged
together with a Printed-Circuit-Board (PCB). Light blue region in
the middle is the chip metallization. In orange is the copper metallization
and in burgundy color is the dielectric of the PCB. b) A bounding
box is defined over the side surfaces of which the drive ports are
defined as rectangles; see the rectangular area in magenta in Fig.
(d). These rectangles should be chosen large enough to enclose all
the fields due to the drive excitations. c) A close-up view of the
drive line of one of the qubits. The copper central trace of the drive
line in the PCB is capacitively coupled to the on-chip readout resonator
(The meandered structures) with a wirebond and a launchpad. (d) The
same picture in (c) with parts of the PCB made transparent to make
the drive port visible (Magenta rectangle on the side surface of the
bounding box). The drive port is usually defined as a wave port (to
which it is assumed that a constant impedance transmission line is
connected) and the size of the rectangle should be chosen properly
to enclose the fields due to the excitation at the port. e) Close-up
view of qubit pads (two identical rectangles in light blue). Light
gray are wirebonds and the pink region in the qubit pocket is the
upper surface of the substrate underneath the chip metallization.
f) Qubit port is the small square shown in magenta defined between
the leads connecting the qubit pads (large light blue regions) to
the Josepson junction(not shown).}
\end{figure}

\twocolumngrid
}

\clearpage{}

\subsection{\label{sub:Definition-of-the-ports-in-HFSS}Defining the Qubit Ports
and the Drive Ports in the 3D Finite-Element Electromagnetic Simulators}

In the main text we described in words how to define the qubit ports
and the drive ports. In this appendix we illustrate the definition
of the ports with the help of the 3D model of a 7-Qubit device in
HFSS \cite{HFSS} as shown in Fig. \eqref{fig:Port-Definitions} (HFSS
is a high-frequency finite-element electromagnetics simulator). The
device consists of a quantum chip (shown in light blue in the middle)
packaged together with a PCB (Printed Circuit Board) supporting transmission
lines carrying the drive and readout signals to/from the chip. The
metallization of the PCB is shown in orange and the dielectric of
the PCB is shown in burgundy color in Fig. \eqref{fig:Port-Definitions}.
For the definition of the drive ports we choose a bounding box enclosing
the quantum chip and some part of the PCB. The boundaries of the box
should be chosen far enough from the chip. As we stated in the main
text the exact position of this boundary can be determined with a
TDR (Time-Domain Reflectometry) experiment/simulation. Ideally we
would like to put the boundary at the location where signals traveling
in the transmission lines of the PCB start to see a change in the
constant impedance of the transmission lines. This happens where the
signals enter the discontinuity region between the PCB and the chip.
The drive ports are usually defined as wave ports in HFSS to which
it is assumed that a constant impedance transmission line is connected.
An example of a drive port is shown in the sub-figure (d) in Fig.
\eqref{fig:Port-Definitions} as the magenta rectangle on one of the
side surfaces of the bounding box shown in sub-figure (b) in Fig.
\eqref{fig:Port-Definitions}.

Qubit Ports are defined as lumped ports in HFSS. This is shown in
sub-figures (e) and (f) in Fig. \eqref{fig:Port-Definitions}. The
qubit port is the small magenta square shown in sub-figure (f) in
Fig. \eqref{fig:Port-Definitions}. HFSS puts a differential excitation
between the edges of this square touching the junction terminals.

\clearpage{}

\subsection{Expansion of the junction potentials\label{sub:Expansion-of-the-junction-potentials}}

Qubit anharmonicities and dispersive shifts between the modes are
obtained after including the nonlinear terms in the junction potentials.
For this we use the normal ordered expansion as given in Eq. (16)
of \cite{BBQ-Yale}:
\begin{equation}
H=H_{0}+H_{\gamma}+H_{\beta}+\mathcal{O}(\hat{\varphi}_{J}^{6})\label{eq:Yale-normal-ordered-expansion}
\end{equation}
with

\begin{widetext}

\begin{eqnarray}
H_{\gamma} & = & -\underset{pp'}{\sum}\gamma_{pp'}(2\hat{a}_{p}^{\dagger}a_{p'}+\hat{a}_{p}\hat{a}_{p'}+\hat{a}_{p}^{\dagger}\hat{a}_{p'}^{\dagger})\label{eq:H-omega}\\
H_{\beta} & = & \underset{pp'qq'}{-\sum}\beta_{pp'qq'}(6\hat{a}_{p}^{\dagger}\hat{a}_{p'}^{\dagger}\hat{a}_{q}\hat{a}_{q'}+4\hat{a}_{p}^{\dagger}\hat{a}_{p'}^{\dagger}\hat{a}_{q}^{\dagger}\hat{a}_{q'}+4\hat{a}_{p}^{\dagger}\hat{a}_{p'}\hat{a}_{q}\hat{a}_{q'}+\hat{a}_{p}\hat{a}_{p'}\hat{a}_{q}\hat{a}_{q'}+\hat{a}_{p}^{\dagger}\hat{a}_{p'}^{\dagger}\hat{a}_{q}^{\dagger}\hat{a}_{q'}^{\dagger})\label{eq:H-beta}
\end{eqnarray}
\end{widetext}where $H_{0}$ is the linearized part of the Hamiltonian
obtained after replacing the junctions with linear inductors, $p$,
$p'$, $q$, $q'$ are the labels of the harmonic modes in the basis
defined by $H_{0}$ and $\hat{a}_{p}(\hat{a}_{p}^{\dagger})$ is the
annihilation(creation) operator of the mode $p$. This expansion was
originally done in \cite{BBQ-Yale} in a diagonal frame whereas here
we will expand in the block-diagonalized frame corresponding to the
matrix $\widetilde{\mathbf{M}}_{1}$ in Eq. \eqref{eq:block-diagonal-M1}.
That is the linearized Hamiltonian $H_{0}$ in our case is the linear
part of the Hamiltonian given in Eq. \eqref{eq:block-diagonal-hamiltonian}

\begin{equation}
H_{0}=\frac{1}{2}\boldsymbol{q}^{T}\boldsymbol{q}+\frac{1}{2}\boldsymbol{\phi}^{T}\widetilde{\mathbf{M}}_{1}\boldsymbol{\phi}
\end{equation}
In that frame the capacitance matrix is unity and the coordinate vector
holds the flux variables $\boldsymbol{\phi}=(\phi_{1}\ldots\phi_{N+M})$.
The first $N$ coordinates correspond to qubit modes and the last
$M$ coordinates correspond to the resonator modes(or internal modes).
$\boldsymbol{q}$ is the vector of momenta conjugate to coordinates
$\boldsymbol{\phi}$ . The flux operators $\boldsymbol{\phi}$ of
the modes in the final frame can be related to the fluxes $\mathbf{\Phi}$
in the initial frame by the total coordinate transformation

\begin{equation}
\mathbf{\Phi}=\left(\begin{array}{c}
\mathbf{\Phi}_{J}\\
\mathbf{\Phi}_{R}
\end{array}\right)=\left(\begin{array}{cc}
\mathbf{C}_{0}^{-1/2} & \mathbf{0}\\
\mathbf{0} & \mathbf{1}
\end{array}\right)\boldsymbol{\alpha}\boldsymbol{\phi}
\end{equation}
where $\boldsymbol{\alpha}=\mathbf{T}\exp(\mathbf{S})$; matrices
$\mathbf{C}_{0}$, $\mathbf{T}$, $\mathbf{S}$ are defined in the
text in Eqs. \eqref{eq:Capacitance-matrix}, \eqref{eq:T-transformation}
and \eqref{eq:block-diagonal-M1}, respectively. In particular

\begin{eqnarray}
\mathbf{\Phi}_{J} & = & \left(\begin{array}{cc}
\mathbf{C}_{0}^{-1/2} & \mathbf{0}_{N\times M}\end{array}\right)\boldsymbol{\alpha}\boldsymbol{\phi}
\end{eqnarray}
where $\mathbf{\Phi}_{J}=(\Phi_{J_{1}}\ldots\Phi_{J_{N}})^{T}$ is
the vector of fluxes across the Josephson junctions. Hence

\begin{equation}
\Phi_{J_{i}}=\frac{1}{\sqrt{C_{i}}}\left[\stackrel[j=1]{N}{\sum}\alpha_{ij}\phi_{j}+\stackrel[k=1]{M}{\sum}\alpha_{i,k+N}\phi_{k}\right]
\end{equation}
where indices $i$, $j$ label qubit modes and $k$ labels resonator
modes with $1\leq i,j\leq N$ and $1\leq k\leq M$. The $(N+M)\times(N+M)$
matrix $\boldsymbol{\alpha}$ has the entries

\begin{eqnarray}
\alpha_{ii} & = & 1-\mathrm{Im}[Z_{ii}^{(AC)}(\omega_{i})]/Z_{i}\label{eq:alpha-ii-exp}\\
\alpha_{ij} & = & -\mathrm{Im}[Z_{ij}(\omega_{i})]/Z_{i}\\
\alpha_{i,k+N} & = & r_{ki}C_{i}^{1/2}\left(\frac{\omega_{R_{k}}^{2}}{\omega_{R_{k}}^{2}-\omega_{i}^{2}}\right)\label{eq:alpha-ik-exp}
\end{eqnarray}
In the dispersive regime we have $\left|\mathrm{Im}[Z_{ii}^{(AC)}(\omega_{i})]/Z_{i}\right|\ll1$
hence $\alpha_{ii}\simeq1$, $\left|\alpha_{ij}\right|=\left|\mathrm{Im}[Z_{ij}(\omega_{i})]/Z_{i}\right|\ll1$
for $1\leq i,j\leq N$ and $\left|\alpha_{i,k+N}\right|=\left|r_{ki}C_{i}^{1/2}\left(\frac{\omega_{R_{k}}^{2}}{\omega_{R_{k}}^{2}-\omega_{i}^{2}}\right)\right|\ll1$
for $1\leq k\leq M$ hence we can treat $\alpha_{ij}$ and $\alpha_{i,k+N}$'s
as small parameters. $\mathrm{Im}[Z_{ii}^{(AC)}(\omega_{i})]=\stackrel[k=1]{M}{\sum}\frac{\left[A_{k}\right]_{ii}\omega_{i}}{\omega_{R_{k}}^{2}-\omega_{i}^{2}}$
is the AC part of $\mathrm{Im}[Z_{ii}(\omega_{i})]$.

Similarly resonator fluxes $\mathbf{\Phi}_{R}=(\Phi_{R_{1}},\ldots,\Phi_{R_{M}})$
in the initial frame can also be related to the flux coordinates $\boldsymbol{\phi}$
in the final frame

\begin{equation}
\Phi_{R_{k}}=\stackrel[i=1]{N}{\sum}\alpha_{k+N,i}\phi_{i}+\phi_{k}\label{eq:resonator-fluxes}
\end{equation}
where

\begin{eqnarray}
\alpha_{k+N,i} & = & r_{ki}C_{i}^{1/2}\left(\frac{\omega_{i}^{2}}{\omega_{i}^{2}-\omega_{R_{k}}^{2}}\right)
\end{eqnarray}
and $\alpha_{k+N,k+N}=1$ for $1\leq k\leq M$ by Eq. \eqref{eq:resonator-fluxes}.

The expression for the coefficients $\beta_{pp'qq'}$ is given in
\cite{BBQ-Yale} as

\begin{equation}
\beta_{pp'qq'}=\stackrel[s=1]{N}{\sum}\frac{e^{2}}{24L_{J}^{(s)}}\xi_{sp}\xi_{sp'}\xi_{sq}\xi_{sq'}
\end{equation}
where $L_{J}^{(s)}$ is the inductance of the $s^{th}$ junction.
In our case $\xi_{sp}=\alpha_{sp}/\sqrt{\omega_{p}}$ after the introduction
of mode operators as $\hat{\phi}_{p}=\sqrt{\frac{\hbar Z_{p}}{2}}(\hat{a}_{p}+\hat{a}_{p}^{\dagger})$
in the final frame with characteristic impedance $Z_{p}=1/\omega_{p}$
one gets

\begin{equation}
\beta_{pp'qq'}=\stackrel[s=1]{N}{\sum}\frac{E_{C}^{(s)}}{12}\omega_{J_{s}}^{2}\left(\omega_{p}\omega_{p'}\omega_{q}\omega_{q'}\right)^{-1/2}\alpha_{sp}\alpha{}_{sp'}\alpha{}_{sq}\alpha_{sq'}\label{eq:beta-expression}
\end{equation}
where $E_{C}^{(s)}=\frac{e^{2}}{2C_{s}}$ and $\omega_{J_{s}}=1/\sqrt{L_{J_{s}}C_{s}}$.

$\gamma_{pp'}$ is given in \cite{BBQ-Yale} as

\begin{equation}
\gamma_{pp'}=6\stackrel[q=1]{N+M}{\sum}\beta_{qqpp'}
\end{equation}
We observe that $\gamma_{pp}$ is of order $E_{C}^{(s)}$. Since $E_{C}^{(s)}$
is already small compared to qubit frequencies we will be only interested
in the first order expansion of $\gamma_{pp'}$ in the small parameters
$\alpha_{ij}$'s and $\alpha_{i,k+N}$'s. Then we can write the diagonal
entries as

\begin{eqnarray}
\gamma_{ii} & \simeq & 6\beta_{iiii}\nonumber \\
 & \simeq & \frac{E_{C}^{(i)}\omega_{J_{i}}^{2}}{2\omega_{i}^{2}}\alpha_{ii}^{2}
\end{eqnarray}
The off-diagonal entry $\gamma_{ij}$ between qubit modes $i$ and
$j$ is

\begin{eqnarray}
\gamma_{ij} & \simeq & 6(\beta_{iiij}+\beta_{jjij})\nonumber \\
 & \simeq & \frac{E_{C}^{(i)}\omega_{J_{i}}^{2}}{2\omega_{i}}\frac{\alpha_{ii}\alpha_{ij}}{\sqrt{\omega_{i}\omega_{j}}}+\frac{E_{C}^{(j)}\omega_{J_{j}}^{2}}{2\omega_{j}}\frac{\alpha_{jj}\alpha_{ji}}{\sqrt{\omega_{i}\omega_{j}}}
\end{eqnarray}
The off-diagonal entry $\gamma_{ik}$ between the qubit mode $i$
and the resonator mode $k$ is

\begin{eqnarray}
\gamma_{ik} & \simeq & 6\beta_{iiik}\nonumber \\
 & \simeq & \frac{E_{C}^{(i)}\omega_{J_{i}}^{2}}{2\omega_{i}}\frac{\alpha_{ii}\alpha_{i,k+N}}{\sqrt{\omega_{i}\omega_{R_{k}}}}
\end{eqnarray}
And the diagonal resonator entries $\gamma_{kk}\simeq0$ to first
order in $\alpha_{ij}$ and $\alpha_{i,k+N}$'s. Hence we can write

\begin{equation}
\boldsymbol{\gamma}=\sqrt{\mathbf{z}}\boldsymbol{\alpha}^{T}\boldsymbol{\Lambda}\mathbf{\boldsymbol{\alpha}}\sqrt{\mathbf{z}}\label{eq:gamma-matrix}
\end{equation}
where $\boldsymbol{\Lambda}$ is the $(N+M)\times(N+M)$ diagonal
matrix with entries $\frac{E_{C}^{(i)}\omega_{J_{i}}^{2}}{2\omega_{i}}$'s
for $1\leq i\leq N$ and zero otherwise, that is

\begin{equation}
\boldsymbol{\Lambda}=\left(\begin{array}{cccc}
\frac{E_{C}^{(1)}\omega_{J_{1}}^{2}}{2\omega_{1}} &  &  & \mathbf{0}\\
 & \ddots\\
 &  & \frac{E_{C}^{(N)}\omega_{J_{N}}^{2}}{2\omega_{N}}\\
\mathbf{0} &  &  & \mathbf{0}_{M\times M}
\end{array}\right)
\end{equation}
and $\sqrt{\mathbf{z}}$ is the diagonal matrix holding the square
roots of the characteristic impedances $Z_{i}=1/\omega_{i}$ of the
modes in the final frame

\begin{equation}
\sqrt{\mathbf{z}}=\left(\begin{array}{ccc}
1/\sqrt{\omega_{1}}\\
 & \ddots\\
 &  & 1/\sqrt{\omega_{N+M}}
\end{array}\right)
\end{equation}

$H_{\gamma}$ in Eq. \eqref{eq:H-omega} can then be written as

\begin{equation}
H_{\gamma}=-\underset{pp'}{\sum}\gamma_{pp'}(2\hat{a}_{p}^{\dagger}a_{p'}+\hat{a}_{p}\hat{a}_{p'}+\hat{a}_{p}^{\dagger}\hat{a}_{p'}^{\dagger})=-\frac{1}{2}\hat{\boldsymbol{\phi}}^{T}\boldsymbol{\gamma}'\hat{\boldsymbol{\phi}}
\end{equation}
where

\begin{equation}
\boldsymbol{\gamma}'=\left(\frac{4}{\hbar}\right)\sqrt{\mathbf{z}}^{-1}\boldsymbol{\gamma}\sqrt{\mathbf{z}}^{-1}
\end{equation}
Using Eq. \eqref{eq:gamma-matrix}

\begin{equation}
\boldsymbol{\gamma}'=\left(\frac{4}{\hbar}\right)\mathbf{\boldsymbol{\alpha}}^{T}\boldsymbol{\Lambda}\boldsymbol{\mathbf{\alpha}}
\end{equation}

Then one can show that $H_{\gamma}$ when transformed back to the
original frame becomes 

\begin{eqnarray}
H_{\gamma} & = & -\frac{1}{2}\boldsymbol{\phi}^{T}\boldsymbol{\gamma}'\boldsymbol{\phi}\nonumber \\
 & = & -\frac{2}{\hbar}\boldsymbol{\phi}^{T}\mathbf{\boldsymbol{\alpha}}^{T}\boldsymbol{\varLambda}\mathbf{\boldsymbol{\alpha}}\boldsymbol{\phi}\nonumber \\
 & = & -\frac{2}{\hbar}\mathbf{\mathbf{\Phi}}^{T}\left(\begin{array}{cc}
\mathbf{C}_{0}^{1/2} & \mathbf{0}\\
\mathbf{0} & \mathbf{1}
\end{array}\right)\boldsymbol{\varLambda}\left(\begin{array}{cc}
\mathbf{C}_{0}^{1/2} & \mathbf{0}\\
\mathbf{0} & \mathbf{1}
\end{array}\right)\mathbf{\mathbf{\Phi}}\nonumber \\
 & = & -\frac{1}{2}\mathbf{\Phi}_{J}^{T}\mathbf{L}_{0}^{-1}\mathbf{\Phi}_{J}
\end{eqnarray}
where $\mathbf{L}_{0}$ is a diagonal inductance matrix

\begin{equation}
\mathbf{L}_{0}=\mathbf{L}_{J}\left(\begin{array}{ccc}
\frac{\hbar\omega_{1}}{2E_{C}^{(1)}} &  & \boldsymbol{0}\\
 & \ddots\\
\boldsymbol{0} &  & \frac{\hbar\omega_{N}}{2E_{C}^{(N)}}
\end{array}\right)
\end{equation}

Now if we write the initial Hamiltonian $H_{0}$ by adding and subtracting
the term $H_{\gamma}$ as

\begin{eqnarray}
H_{0} & = & H_{0}+H_{\gamma}-H_{\gamma}\nonumber \\
 & = & H'_{0}-H_{\gamma}
\end{eqnarray}
where $H'_{0}=H_{0}+H_{\gamma}$. So instead of starting our treatment
with $H_{0}$ if we start with an initial linear Hamiltonian $H'_{0}$
we would cancel out the term $H_{\gamma}$ that is generated by the
non-linearities. This requires an update of the junction inductances
in the initial frame as follows

\begin{eqnarray}
\mathbf{L}_{J}^{-1} & \rightarrow & \mathbf{L}_{J}^{-1}-\mathbf{L}_{0}^{-1}
\end{eqnarray}
That is

\begin{equation}
L_{J_{i}}^{-1}\rightarrow L_{J_{i}}^{-1}\left(1-\frac{2E_{C}^{(i)}}{\hbar\omega_{i}}\right)
\end{equation}
Hence we can write the equation for $\omega_{i}$

\begin{equation}
\omega_{i}^{2}=\omega_{J_{i}}^{2}\left(1-\frac{2E_{C}^{(i)}}{\hbar\omega_{i}}\right)
\end{equation}
or if we put $r=\frac{E_{C}^{(i)}}{\hbar\omega_{J_{i}}}$ and $x=\omega_{i}/\omega_{j}$

\begin{equation}
x^{2}=\left(1-2r/x\right)
\end{equation}
In the limit of small anharmonicities $r\ll1$ the solution is $x=1-r/(1-r)$
or

\begin{equation}
\omega_{i}=\omega_{J_{i}}-\frac{E_{C}^{(i)}/\hbar}{1-E_{C}^{(i)}/(\hbar\omega_{J_{i}})}\label{eq:qubit-freq-solution}
\end{equation}


\begin{thebibliography}{10}
\bibitem{Koch-Transmon}J. Koch, T. M. Yu, J. Gambetta, A. A. Houck,
D. I. Schuster, J. Majer, A. Blais, M. H. Devoret, S. M. Girvin, and
R. J. Schoelkopf, Phys. Rev. A 76, 042319 (2007).

\bibitem{Zombie-paper-Gambetta}Gambetta J. M., Murray C. E., Fung
Y. K. K., McClure D. T., Dial O., Shanks W., Sleight J. and Steffen
M., IEEE Trans. Appl. Supercond. 27 1700205 (2016).

\bibitem{Xmon}R. Barends, J. Kelly, A. Megrant, D. Sank, E. Jeffrey,
Y. Chen, Y. Yin, B. Chiaro, J. Mutus, C. Neill, P. O\textquoteright Malley,
P. Roushan, J. Wenner, T. C. White, A. N. Cleland, and John M. Martinis
Phys. Rev. Lett. 111, 080502 (2013).

\bibitem{Dicarlo}D. Ristè, C.C. Bultink, M.J. Tiggelman, R.N. Schouten,
K.W. Lehnert, and L. DiCarlo, Nature Communications 4, 1913 (2013).

\bibitem{Martinis-Nature}R. Barends, J. Kelly, A. Megrant, A. Veitia,
D. Sank, E. Jeffrey, T. C. White, J. Mutus, A. G. Fowler, B. Campbell,
Y. Chen, Z. Chen, B. Chiaro, A. Dunsworth, C. Neill, P. O'Malley,
P. Roushan, A. Vainsencher, J. Wenner, A. N. Korotkov, A. N. Cleland,
John M. Martinis Nature 508, 500-503 (2014).

\bibitem{Sarah-Single-Qubit}Sarah Sheldon, Lev S. Bishop, Easwar
Magesan, Stefan Filipp, Jerry M. Chow, and Jay M. Gambetta Phys. Rev.
A 93, 012301 (2016).

\bibitem{Sarah-CR}Sarah Sheldon, Easwar Magesan, Jerry M. Chow, and
Jay M. Gambetta Phys. Rev. A 93, 060302(R) (2016).

\bibitem{Maika-Parity}Maika Takita, A.\LyXThinSpace D. Córcoles,
Easwar Magesan, Baleegh Abdo, Markus Brink, Andrew Cross, Jerry M.
Chow, and Jay M. Gambetta Phys. Rev. Lett. 117, 210505 (2016).

\bibitem{Nick}Nicholas T. Bronn, Vivekananda P. Adiga, Salvatore
B. Olivadese, Xian Wu, Jerry M. Chow, David P. Pappas, arXiv:1709.02402.

\bibitem{Jaynes-Cummings}E. Jaynes and F. Cummings, Proceedings of
the IEEE 51, 89 (1963).

\bibitem{Blais-xQED}Alexandre Blais, Ren-Shou Huang, Andreas Wallraff,
S. M. Girvin, and R. J. Schoelkopf Phys. Rev. A 69, 062320 (2004).

\bibitem{Wallraff}A. Wallraff, D. I. Schuster, A. Blais, L. Frunzio,
R.- S. Huang, J. Majer, S. Kumar, S. M. Girvin, and R. J. Schoelkopf,
Nature (London) 431, 162 (2004).

\bibitem{Controlling Spontaneous Emission - Houck}A. A. Houck, J.
A. Schreier, B. R. Johnson, J. M. Chow, Jens Koch, J. M. Gambetta,
D. I. Schuster, L. Frunzio, M. H. Devoret, S. M. Girvin, and R. J.
Schoelkopf, Phys. Rev. Lett. 101, 080502, (2008).

\bibitem{Bourassa-Multi-Mode-circuit-QED}Jerome Bourassa, Jay M.
Gambetta, and Alexandre Blais, \textquotedblleft Multi-mode circuit
quantum electrodynamics,\textquotedblright , Abstract Y29.00005, APS
March Meeting, Dallas, 2011.

\bibitem{Gely-Adrian-Solano}Mario F. Gely, Adrian Parra-Rodriguez,
Daniel Bothner, Ya. M. Blanter, Sal J. Bosman, Enrique Solano, Gary
A. Steele, Phys. Rev. B 95, 245115 (2017).

\bibitem{Adrian-Long}A. Parra-Rodriguez, E. Rico, E. Solano, I. L.
Egusquiza, arXiv:1711.08817.

\bibitem{Devoret-Les-Houches}Michel H. Devoret, in Quantum fluctuations,
Les Houches, Elsevier, Amsterdam, (1997).

\bibitem{BKD}G. Burkard, R. H. Koch, and D. P. DiVincenzo, Phys.
Rev. B 69, 064503 (2004).

\bibitem{Burkard}Guido Burkard, Phys. Rev. B 71, 144511, (2005).

\bibitem{Foster}Foster, R. M., \textquotedblleft A reactance theorem\textquotedblright ,
Bell Systems Technical Journal, vol.3, no. 2, pp. 259\textendash 267,
November 1924.

\bibitem{Brune}O. Brune, Synthesis of a finite two-terminal network
whose driving-point impedance is a prescribed function of frequency,
Doctoral thesis, MIT, 1931.

\bibitem{Newcomb}Robert W. Newcomb, \emph{Linear Multiport Synthesis},
McGraw-Hill, 1966.

\bibitem{BBQ-Yale}S. E. Nigg, H. Paik, B. Vlastakis, G. Kirchmair,
S. Shankar, L. Frunzio, M. H. Devoret, R. J. Schoelkopf, and S. M.
Girvin, Phys. Rev. Lett. 108, 240502 (2012).

\bibitem{Brune-Quantization}Firat Solgun, David W. Abraham, and David
P. DiVincenzo, Phys. Rev. B 90, 134504, (2014).

\bibitem{Solgun}Firat Solgun and David P. DiVincenzo, Annals of Physics,
Vol. 361, pp. 605-669, October 2015.

\bibitem{DDV}Although \cite{Newcomb} considers general full-rank
matrices which would correspond to having multiple degenerate internal
modes we argue that in real physical systems small couplings will
remove any such degeneracy.

\bibitem{Pozar}David M. Pozar, \emph{Microwave Engineering}, 3rd
ed., John Wiley \& Sons, 2005.

\bibitem{Brito}D. P. DiVincenzo, Frederico Brito, and Roger H. Koch,
Phys. Rev. B 74, 014514 (2006).

\bibitem{Winkler}Roland Winkler, \emph{Spin-Orbit Coupling Effects
in Two-Dimensional Electron and Hole Systems}, STMP 191, 201-205,
Springer-Verlag Berlin Heidelberg, 2003.

\bibitem{Jay-Juelich}\textcolor{black}{Jay M. Gambetta, Lecture Notes
of the $44^{th}$ IFF Spring School ``Quantum Information Processing''
(Forschungszentrum Jülich, 2013).}

\bibitem{Wirebond-Crosstalk-Martinis}J. Wenner, M. Neeley, Radoslaw
C. Bialczak, M. Lenander, Erik Lucero, A. D. O'Connell, D. Sank, H.
Wang, M. Weides, A. N. Cleland, John M Martinis Superconductor Science
and Technology 24, 065001 (2011).

\bibitem{IBM-Q-Experience}IBM Q Experience, https://quantumexperience.ng.bluemix.net/.

\bibitem{HFSS}Ansys HFSS (High Frequency Structural Simulator), http://www.ansys.com.\end{thebibliography}
\end{document}